\documentclass[11pt,preprint,authoryear]{elsarticle}
\usepackage{natbib}
\usepackage{lineno}
\biboptions{authoryear}
\usepackage{geometry}                
\usepackage{graphicx}
\usepackage{amssymb}
\usepackage{epstopdf}
\usepackage{setspace}
\usepackage{subfigure}
\newcommand{\myleft}{16mm}
\newcommand{\mybottom}{50mm}
\newcommand{\myright}{25mm}
\newcommand{\mytop}{60mm}

\begin{document}

\onehalfspacing

\begin{frontmatter}
\title{Modelling the brightness increase signature due to asteroid collisions}
\author{Ev McLoughlin}
\ead{emcloughlin03@qub.ac.uk}
\author{Alan Fitzsimmons}
\author{Alan McLoughlin}

\address{Astrophysics Research Centre, School of Mathematics and Physics, Queen's University Belfast. University~Rd, Belfast, Northern Ireland, UK, BT7 1NN}

\begin{abstract}
We have developed a model to predict the post-collision brightness increase of sub-catastrophic collisions between asteroids and to evaluate the likelihood of a survey detecting these events. It is based on the cratering scaling laws of \cite{HH2007} and models the ejecta expansion following an impact as occurring in discrete shells each with their own velocity. We estimate the magnitude change between a series of target/impactor pairs, assuming it is given by the increase in reflecting surface area within a photometric aperture due to the resulting
ejecta. As expected the photometric signal increases with impactor size, but we find also that the photometric signature  decreases rapidly as the target asteroid diameter increases, due to gravitational fallback.  We have used the model results to make an estimate of the impactor diameter for the (596) Scheila collision of $D=49-65$m depending on the impactor taxonomy, which is broadly consistent with previous estimates. We varied both the strength regime (highly porous and sand/cohesive soil) and the taxonomic type (S-, C- and D-type) to examine the effect on the magnitude change, finding that it is significant at early stages but has only a small effect on the overall lifetime of the photometric signal. Combining the results of this model with the collision frequency estimates of \cite{Bottke2005}, we find that low-cadence surveys of $\sim$one visit per lunation will  be insensitive to impacts on asteroids with $D<20$km if relying on photometric detections.
\end{abstract}
 

\end{frontmatter}
\section{Introduction}

The main asteroid belt is collisionally dominated with large asteroids' shapes, sizes and surface geology controlled by impacts. Studies of collisions help us to understand the evolution of the shape of the asteroid population and in turn the formation of our Solar system. These studies may involve laboratory experiments, computer modelling or observational programmes.

The evidence for collisions can be seen indirectly in main-belt asteroid families \citep{families}, asteroid satellites and binaries \citep{satellites}. It can also be seen  directly in recently observed collisions \citep{snodgrass, jewitt, stevenson}.
There are three possible  collisions observed to date.
In 2009 the 120 m diameter asteroid P/2010 A2 suffered a collision with a 6-9 m estimated diameter impactor \citep{snodgrass} (but see section 4.4).
In 2010 another asteroid, (596) Scheila (113 km diameter), was hit with a $\sim35$m diameter impactor \cite{jewitt}.
The most recent potential collision involved the object P/2012 F5 (Gibbs), which like others was originally identified as a potential main-belt comet \cite{stevenson}.
Events like the (596) Scheila collision should occur approximately every 5 years and collisions with asteroids $<$10m even more often \citep{bodewits}. 

Several recent surveys are capable of detecting collisions and cratering events. For example, the Canada-France-Hawaii Telescope Legacy Survey was used to search for Main-Belt comets among 25240 objects in 2003-2009 \citep{cfhtls}, the Thousand Asteroid Lightcurve Survey (924 objects) was conducted with the Canada-France-Hawaii Telescope in September 2006 \citep{tals} and the Hawaii Trails project was conducted in 2009 (599 objects) \citep{htrails}. While none of the surveys mentioned above were specifically looking for main belt collisions, the methods used in search for main belt comets would have also revealed any collisional events.
There are also current surveys fully or partly dedicated to discovering Near Earth Asteroids, such as Pan-STARRS 1 \citep{pstarrs1}, the Lincoln Near-Earth Asteroid Research (LINEAR, responsible for discovery of P/2010 A2) \citep{linear}, the Catalina Sky Survey \citep{catalina} and the VST ATLAS survey \citep{atlas} that are all capable of detecting main-belt collisions. 

Much work has been done in modelling the parameters (i.e. shape of debris, brightness, total ejected mass, impactor mass) of known observed collisions \citep{kleyna}, \citep{ishiguro}, \citep{HH2007}, \citep{HH2011} ; and hydrodynamic modelling of generalised collision \citep{benz}. 
This work focuses solely on the magnitude change following an impact as it is most likely to be observable by optical telescopes. Rather than looking at a specific object in the main belt, the described model looks at what would be expected with generic asteroids.

\section{Model description}

\subsection{Cratering physics}

Our model is based on the work by  \cite{HH2007}, who provide a summary of scaling laws that allows calculation of crater size using properties of the target and impactor, based on the results of impact experiments. These laws can also be used to calculate the evolution of the ejecta dispersal and consequently estimate the amount of material ejected and increase in brightness following a collision.
The decrease in magnitude of the target asteroid is going to depend on the amount of material that was ejected and whether it is optically thin or not. 

At high impact speeds, transfer of the energy and momentum of the impactor into the target occurs over area on the order of impactor size, while the resulting crater usually exceeds this size by many times. It is therefore a reasonable approximation to assume that impact occurs as a point source. Using theoretical analyses of mechanics of crater formation, Holsapple and Housen showed that the crater and ejected material characteristics depend on the quantity  $aU^{\mu} \delta^{\nu}$, where $a$ is the radius, $U$ is the normal velocity component of the impactor and $\delta$ is the density of the impactor; $\mu$ and $\nu$ are scaling exponents. 

The scaling exponents depend on the material properties. Theoretical values of $\mu$ range from 1/3 to 2/3 \citep{numuvalues} and are a measure of the energy dissipation by material; a more porous material can dissipate energy more effectively and will have a lower value of this exponent. Experimentally determined values of $\mu$ are $\sim$0.55 for non-porous materials (e.g. rocks and wet soils), 0.41 for moderately porous materials (e.g. sand and cohesive soils) and 0.33 to 0.40 for highly porous materials \citep{numuvalues}. Experimental values for $\nu$ were found to be the same for all materials at around 0.4 \citep{numuvalues}. By selecting appropriate material scaling parameters for a given impact and inserting them into a general expression for the relationship between radii of involved objects and crater size, a reasonably accurate estimate of the crater size (as well as crater formation time and transient crater growth) can be made.

We now summarise how we use the previous studies in our calculations. Consider a spherical, non-rotating asteroid of radius $r$ following an impact from an object with radius $a$ at sub earth point. The general form of equation for crater size $R$ consists of strength and gravity term:

\begin{equation}
R=aK_{1}(\rm{gravity \:term}+\rm{strength\: term})^{-\frac{\mu}{2+\mu}} 
\end{equation} 

where

\begin{equation}
\rm{gravity\: term}= \frac{S_{g}a}{U^{2}}\big( \frac{\rho}{\delta_{grain}} \big)^{ \frac{2\nu}{\mu}}  
\end{equation} 

\begin{equation}
\rm{strength\: term} =   \big(\frac{Y}{\rho U^{2}}\big)^{ \frac{2+\mu}{2} } \big( \frac{\delta_{grain}}{\rho}\big)^{ \frac{\nu(2+\mu)}{\mu}}   
\end{equation}

Here $K_{1}$ is a scaling parameter (1.03, 1.17, 0.725 for sand/cohesive soil, wet soils/rock and highly porous material respectively; \citep{HH2007}), $Y$ is the average strength of the target material, $\rho$ is the grain density of target, $U = 5$ km s$^{-1}$~\citep{impactvel} is the normal velocity component, $\delta_{grain}$ is the grain density of the impactor, $S_{g}$ is the surface gravity of the target asteroid with mass $M$ and radius $r$, calculated as follows:

\begin{equation}
S_{g}=\frac{GM}{ r^{2}} 
\end{equation}

Depending on the asteroid type, different values of bulk density (for calculation of target asteroid mass) and grain density (for calculation of ejected mass) are used. The values and their sources are summarised in Table \ref{tab:CSD}. Bulk densities of C- and S-type asteroids were taken from weighted averages of corresponding subclasses as summarised in Table 3 of \cite{carry}.  Grain densities of C- and S-types are assumed to be the same as their most likely meteorite analogues \citep{porosity}. Density of D-type asteroids is approximated by bulk and grain densities of the Tagish lake meteorite \citep{Dbulkdensity, Dgraindensity}.

\begin{table}[ht] 
\caption{Average bulk and grain density of asteroids depending on taxonomic type.} 
\centering 
\begin{tabular}{c c c } 
\hline\hline 
Asteroid type & Bulk density (kg m$^{-3}$) & Grain density (kg m$^{-3}$).  \\ [0.5ex] 
\hline 
C & 1840 \citep{carry} & 2710 \citep{porosity} \\ 
S & 2640 \citep{carry} & 3700 \citep{porosity} \\ 
D & 1670 \citep{Dbulkdensity} & 2770 \citep{Dgraindensity} \\ [1ex] 
\hline 
\end{tabular} 
\label{tab:CSD} 
\end{table} 

The range of material strengths used is presented in Table~\ref{tab:strength}. The strength value selected in this study was varied for each taxonomic type to explore the relationship between the type, strength and the corresponding magnitude change. 
\begin{table}[ht] 
\caption{Average material strength used in calculations for a selected scaling.} 
\centering 
\begin{tabular}{c c} 
\hline\hline 
Material & Average strength $Y$ (MPa) \\ [0.5ex] 
\hline 
Sand/cohesive soil & 0.18 \citep{H1993} \\ 
Wet soil/rock & 1.14 \citep{H1993} \\ 
Highly porous & 0.001 \citep{HH2007}\\ [1ex] 
\hline 
\end{tabular} 
\label{tab:strength} 
\end{table} 

The crater radius $R$ calculated in this way has a corresponding mass $M_{\rm{crater}}$:

\begin{equation}
M_{\rm{crater}}=k_{\rm{crater}}\rho R^{3}
\end{equation}

where the scaling factor $k_{\rm{crater}}$ is taken to be 0.75 for cohesive soils, 0.8 for wet soils/rocks or 0.4 for highly porous material \citep{HH2011}.

As we are interested in the ejected mass, since it is only that which contributes to the observed magnitude change of the asteroid, the full crater mass will give an overestimate of brightness. The total crater volume is made up of a volume of ejected mass, a volume of the mass that is uplifted near the crater rim and a volume due to compaction. The fraction $k_{\rm{ejecta}}$  of the total crater mass that corresponds to ejected mass is of order 0.2-0.5 \citep{HH2011}. 

\begin{equation}
M_{\rm{ejecta}}=k_{\rm{ejecta}}M_{\rm{crater}}
\end{equation}

Throughout this work we assume $k_{\rm{ejecta}}$ of 0.3 as being most appropriate to asteroids.

\subsection{Velocity shell model}
\begin{figure}[ht!]
     \begin{center}
       \subfigure[Velocity shells.]{
           \label{fig:shell}
           \includegraphics[trim = 70mm 0.1mm 70mm 20mm, clip, width=0.45\textwidth]{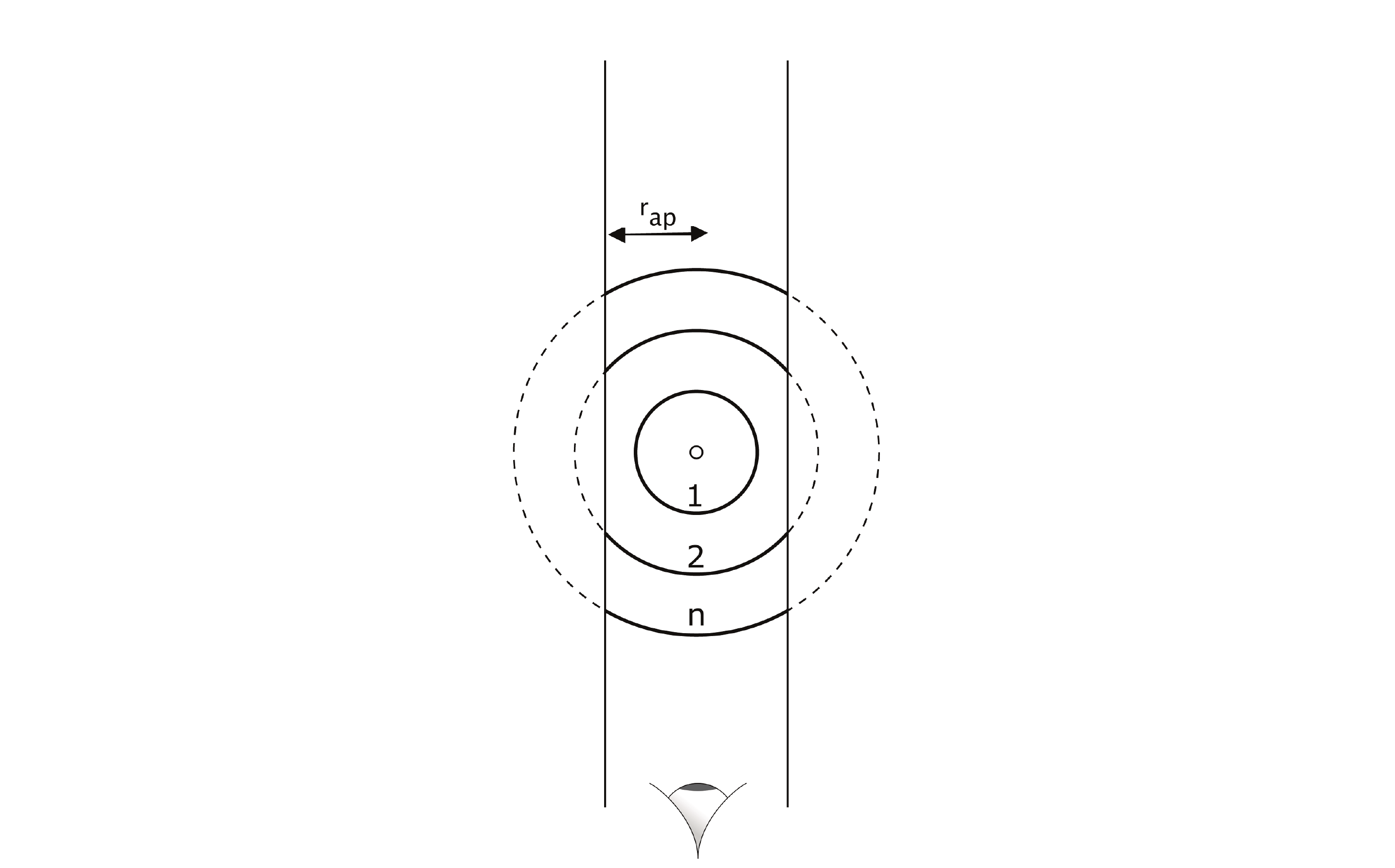}
        }
                \subfigure[Spherical cap.]{
            \label{fig:cap}
            \includegraphics[trim = 50mm 190mm 80mm 30mm, clip, width=0.45\textwidth]{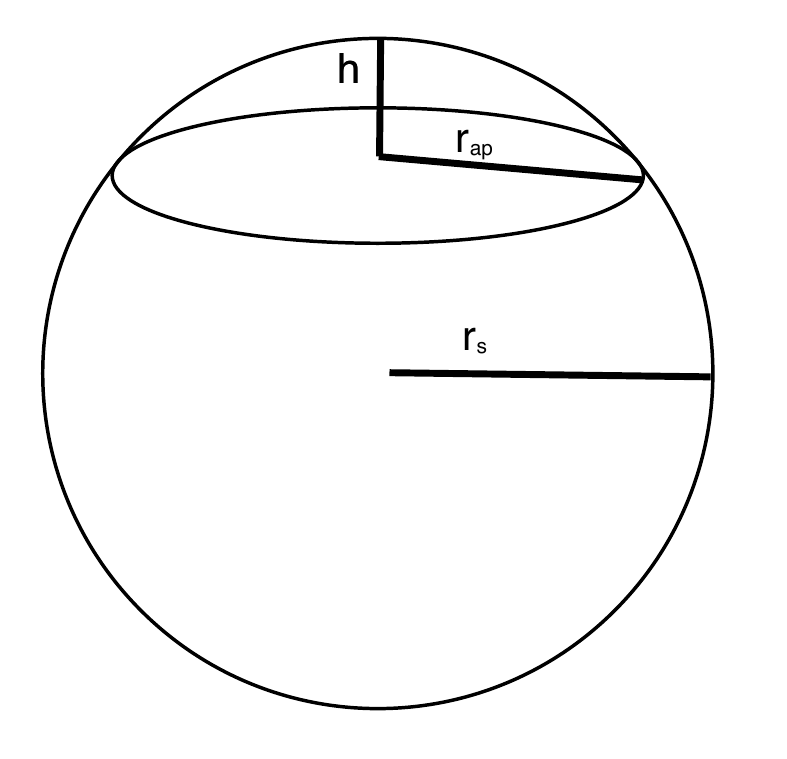}

                      }

    \end{center}
    \caption{Velocity shell model: (a) Ejecta expands spherically outwards from the asteroid. The n$^{th}$ shell expands with velocity $v_{n}$. The portion of each shell that fits into the aperture $r_{ap}$ is marked in solid line. (b) Spherical cap indicates the portion of each shell that fits the aperture.}
         
   \label{fig:velshell}
\end{figure}

We consider the ejecta leaving the asteroid surface after the collision event. For simplicity, we assume that the debris expands spherically outwards from asteroid, the debris with each velocity $v_{n}$ forming a  shell of radius $r_{s}$ (see Figure ~\ref{fig:velshell}). Effects from rotation of the target or impactor are beyond the scope of the current model. Impact experiments show that there is no significant correlation between velocity and mass of the particles \citep{massvel}. Therefore, each velocity shell is taken to have the same particle size distribution described below, in Section \ref{magnitude_change_calculation_section}. As our aim is to model observable brightening from Earth, we assume that the ejecta cloud is centred on the asteroid, as at early epochs the asteroid itself and the ejecta will be unresolved. We also assume that the brightness of the asteroid plus ejecta is measured through an aperture of fixed radius $r_{\rm{ap}}$ and centred on the asteroid. To obtain the amount of material that is visible in the aperture, and consequently the visible increase in asteroid magnitude, we need to look at the fraction of each debris shell that fits completely or partially in the aperture (ejecta visibility fraction $f_{\rm{vis}}$).

\vspace{0.5in}

\vspace{0.5in}
There are three possibilities for an individual shell at a given time $t$:

\begin{enumerate}
  \item the entire shell fits within the aperture and hence $f_{\rm{vis}} = 1$
  \item none of the shell is visible, i.e. the material is not above the surface of the asteroid as it has fallen back, $f_{\rm{vis}} = 0$
  \item part of the shell fits into the aperture ($0<f_{\rm{vis}}<1$) (see Figure ~\ref{fig:velshell}) 
\end{enumerate}

In the latter case, the amount of material visible can be estimated by calculating the ratio of the surface area corresponding to the arc that fits into the aperture (spherical cap) to the total surface area of the sphere. 

\begin{equation}
f_{\rm{vis}} = \frac{2(2 \pi r h)}{4\pi r^{2}}
\end{equation}

Here $h$ is the height of the cap and the extra factor of 2 accounts for near and far side material, $h=r-\sqrt{r^{2}-r_{\rm{ap}}^{2}}$ and $r_{\rm{ap}}$ is the aperture radius that equals the radius of the cap. 
In the situation of no gravity acting on the debris, the radii at each velocity can be described straightforwardly by $r_{s} = vt$, however we need to consider the gravitational case.
For simplicity, we consider a single debris shell of radius $r_{s}$ and velocity $v$ over the asteroid of radius $r$. The acceleration due to the asteroid's gravity that the debris experience is inversely proportional to the square of shell radius ($\frac{1}{r_{s}^{2}}$) and also happens to be the second derivative of radius with respect to time.

\begin{equation}
 \frac{d^{2}r_{s}}{dt^{2}}=Cr_{s}^{-2} 
\end{equation}

$C$ is a constant of proportionality which we calculate by considering the boundary conditions. At the asteroid surface $r_{s} = r$ and the acceleration due to gravity is the  surface gravity $S_{g}$. At the surface the original equation reduces to $Cr^{-2} = S_{g}$, which in turn gives $C= S_{g}r^{2}$. Therefore, the radius of the spherical shell of debris at velocity $v$ is governed by a second order differential equation of the form:

\begin{equation}
 \frac{d^{2}r_{s}}{dt^{2}}=S_{g}r^{2}r_{s}^{-2} 
 \end{equation}

 with the boundary conditions that at $t = 0$, $r_{s} = r$, $\frac{dr_{s}}{dt} = v_{\rm{initial}}$. Solving this equation will give the shell radii at all times $t$. The total mass within the aperture $M_{\rm{vis}}$ is the integral of all velocity shell fractions within the aperture i.e. the shell mass $M_{shell}$ multiplied by $f_{\rm{vis}}$
$$ M_{\rm{vis}} = \int_{v_{\rm{min}}}^{v_{\rm{max}}}M_{\rm{shell}}f_{\rm{vis}}  {\rm{d}}v$$
In the code it is calculated by taking a sum of the individual contributions from each velocity shell, thus
$$M_{\rm{vis}} = \Sigma_{v{\rm{=}}v_{\rm{min}}}^{v_{\rm{max}}}M_{\rm{shell}}f_{\rm{vis}}$$
As the asteroid fragments will be ejected at a continuous range of velocities, increasing the number of shells considered improves the accuracy of calculations.

\subsection{Magnitude change calculation}
\label{magnitude_change_calculation_section}
At this stage we implicitly assume that the scattering function for the asteroid and the ejecta particles is the same and that the ejecta is optically thin throughout.
Assuming all the ejected particles contribute to reflected light with the same albedo as the target asteroid, the total visible surface area after the collision is $A_{c}  = A_{a}+ A_{p}$ where $A_{a}$ is the area of target asteroid, $A_{p}$ is the area of ejected particles. The latter can be calculated assuming that the particle size distribution follows a power law of the following form \citep{ishiguro}:

\begin{equation}
N(a)da = \left\{ \begin{array}{rl}
  N_{0} \big( \frac{a}{a_{0}}\big)^{q}da, &\mbox{ $a_{min}\leq a \leq a_{max}$}  \\
  0, &\mbox{ $a<a_{min}, a>a_{max}$}
       \end{array} \right.
\end{equation}

Here  $a_{min}=0.1$ $\mu$m is chosen to be the minimum particle radius because scattering is inefficient for particle sizes much smaller than the wavelength of the scattered light, $a_{max}$ is the maximum particle radius, $N_{0}$ is the reference dust abundance at reference size $a_{0}$ and $q=3.5$ is the assumed power law index \citep{jewittcent}.
The maximum particle radius $a_{max}$ is calculated using the expression derived from the power-law function of the ejection velocity of dust particles \citep{ishiguro}.

\begin{equation}
a_{max}= \sqrt[k]{ \frac{1}{V_{0}} \sqrt{ \frac{2GM}{r}} } 
\end{equation}

Here $V_{0}$=80 m s$^{-1}$ is the reference ejection velocity of 1 $\mu$m particles,  $k=1/4$ is the power index of size dependence of ejection velocity, $M$ and $r$ are mass and radius of the target respectively.

 After integration we get  cross-sectional area per unit mass of $\sigma_{a}$ (m$^{2}$ kg$^{-1}$), and the total cross-sectional area of particles around the asteroid $A_{p}$ are: 

\begin{equation}
\sigma_{a}=-\frac{3}{4\rho_{grain}}\frac{ a_{max}^{-0.5}-a_{min}^{-0.5} }{ a_{max}^{0.5}-a_{min}^{0.5}} 
\end{equation}

\begin{equation}
A_{p}= \sigma_{a}M_{\rm{vis}}
\end{equation}

The change in magnitude is going to be related to the ratio of the total area after collision  $A_{c}$ and initial target asteroid area $A_{a}$:

\begin{equation}
 \Delta m= - 2.5\log_{10}\frac{A_{c}}{A_{a}} 
\end{equation}

Any ejecta in front of the asteroid will obscure a cross-sectional area equal to its own area. Given that we assume the same scattering function for asteroid and ejecta, this means ejecta directly in front of the asteroid does not contribute to the net brightness. Any ejecta behind the asteroid is clearly not visible. Therefore all ejected material along the line of sight of the asteroid should be ignored for brightness calculations. This is implemented by the subtraction of a pair of spherical caps with radius equal to that of the asteroid.

\section{Modelling results}

\subsection{Model setup}

We have looked at three taxonomic types that are common in the main belt  - C, S and D types \citep{taxtypes}. The differences in the types are accounted for using their corresponding grain and bulk densities during calculation of the ejected mass and magnitude change. 
In the simulation we varied the taxonomic type (C, S, D) and strength (sand/cohesive soil and highly porous) value to see if any patterns emerge. We have used a 50 by 50 point grid of target and impactor radii, ranging from 1-100 km and 1-100 m respectively and logarithmically spaced, and 500 velocity shells linearly spaced over interval $10^{-2}$ to $10^{2}$ m s$^{-1}$. The lowest initial velocity has to be above zero for the purposes of calculations  and  the amount of material being ejected at and beyond $10^{2}$ m s$^{-1}$ is a very small fraction of the total ejected mass and therefore considered negligible, so we take $10^2$ m s$^{-1}$ as a upper limit of the initial velocity range. Given that there is a range of target and impactor pairs that would result in catastrophic disruption (i.e. a dispersive shattering event leaving no remnant larger than 1/2 of the original mass of target asteroid \citep{greenberg2}) rather than in sub catastrophic collision, the results of our model for that region will not be physical. We have estimated the location of the disruption region by calculating the approximate impactor radius necessary for catastrophic disruption of a basalt target assuming average density of 2500 kg m$^{-3}$ and impact velocity of 5 km s$^{-1}$ \citep{benz}. This region is marked in black on the figures.
The code for this model was written in Python 2.7 (includes NumPy and SciPy packages), plotting of the output was performed in MatLab. The code requires a separate run for each pair of values for the taxonomic type and strength regime (i.e. C-type in sand/cohesive soil and C-type in highly porous regime would be two different code runs). Each run from impact until 7 days after impact with the above parameters takes approximately 2 hours on a single core of an Intel Core i5 2410M processor running at 2.3 GHz.

\subsection{Model Results}

\begin{figure}[htp]
     \begin{center}
        \subfigure[highly porous regime, 1 day]{
            \label{fig:CHP24}
            \includegraphics[trim = {\myleft} {\mybottom} {\myright} {\mytop}, clip, width=0.45\textwidth]{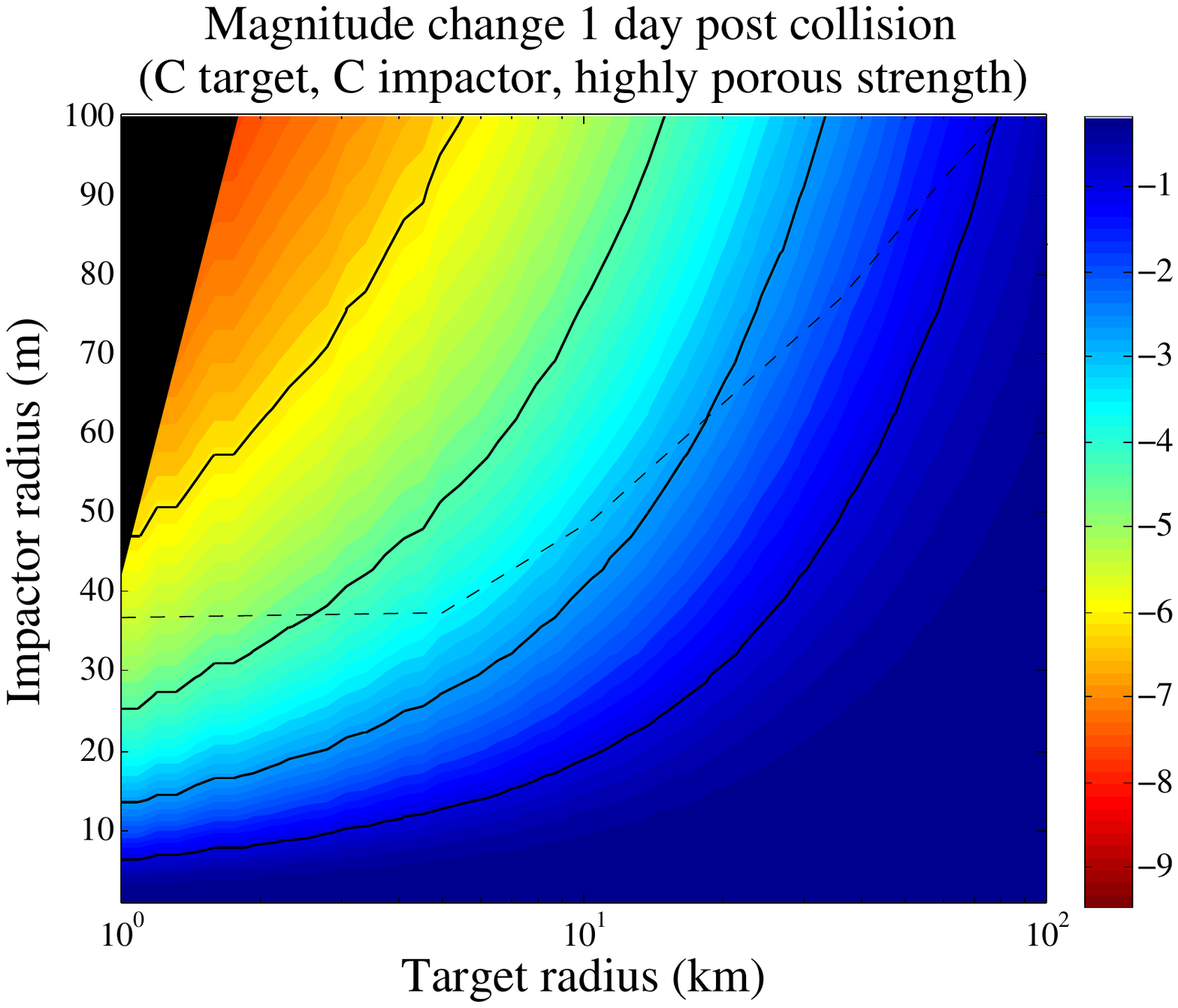}
        }
        \subfigure[highly porous regime, 7 days]{
           \label{fig:CHP1w}
           \includegraphics[trim = {\myleft} {\mybottom} {\myright} {\mytop}, clip,width=0.45\textwidth]{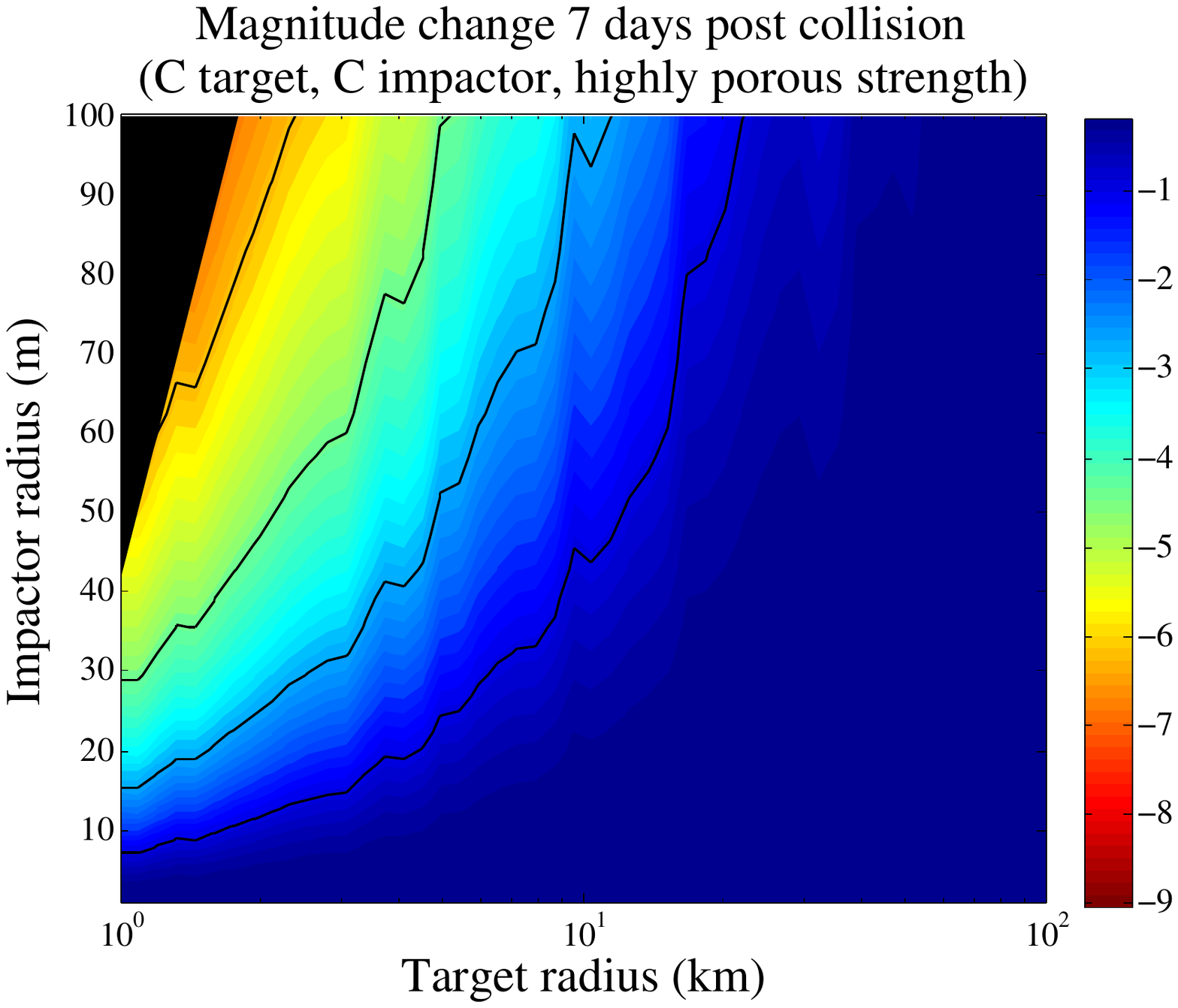}
       }
              \subfigure[sand/cohesive soil regime, 1 day]{
            \label{fig:CSAND24}
            \includegraphics[trim = {\myleft} {\mybottom} {\myright} {\mytop}, clip, width=0.45\textwidth]{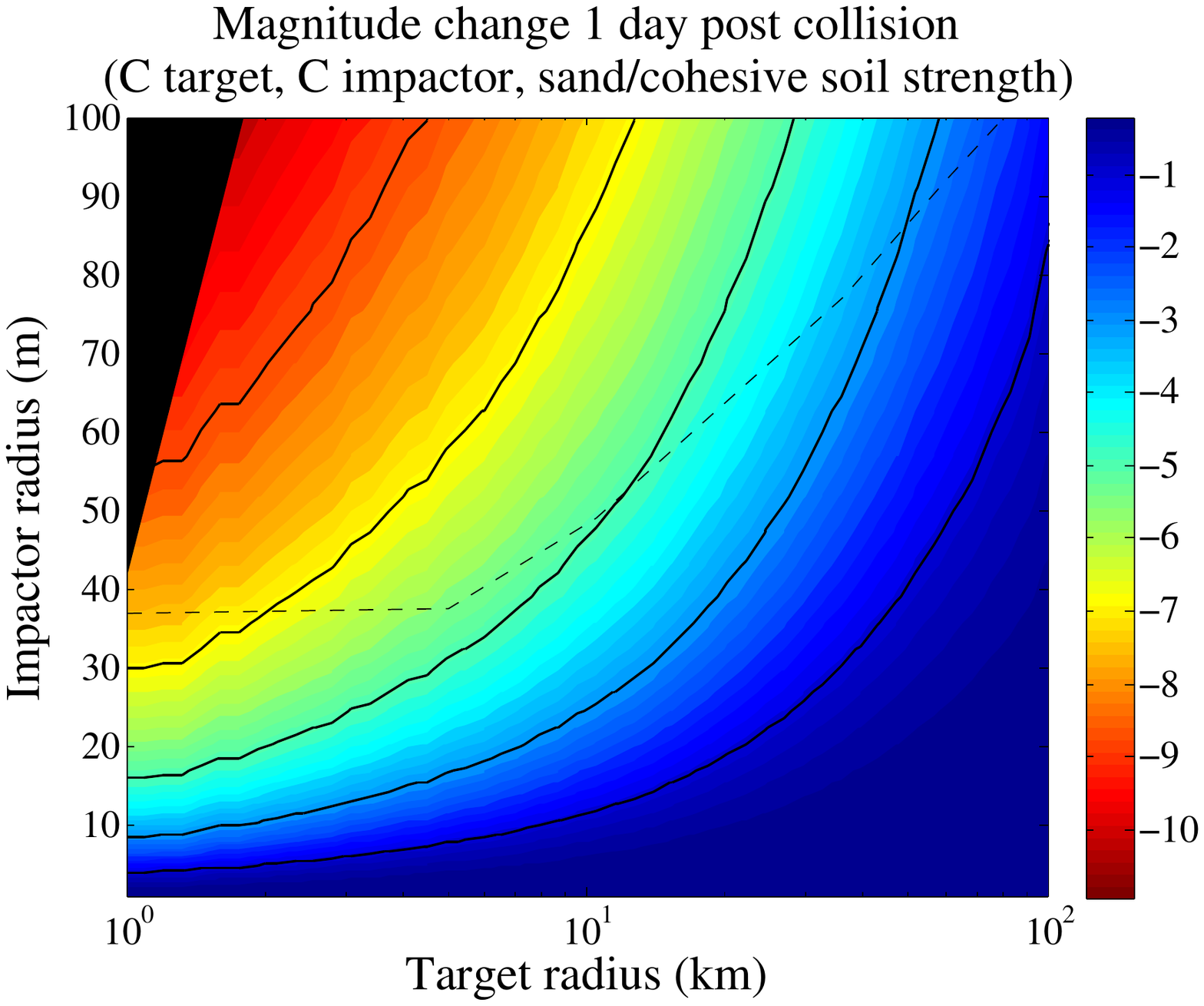}
        }
        \subfigure[sand/cohesive soil regime, 7 days]{
           \label{fig:CSAND1w}
           \includegraphics[trim = {\myleft} {\mybottom} {\myright} {\mytop}, clip,width=0.45\textwidth]{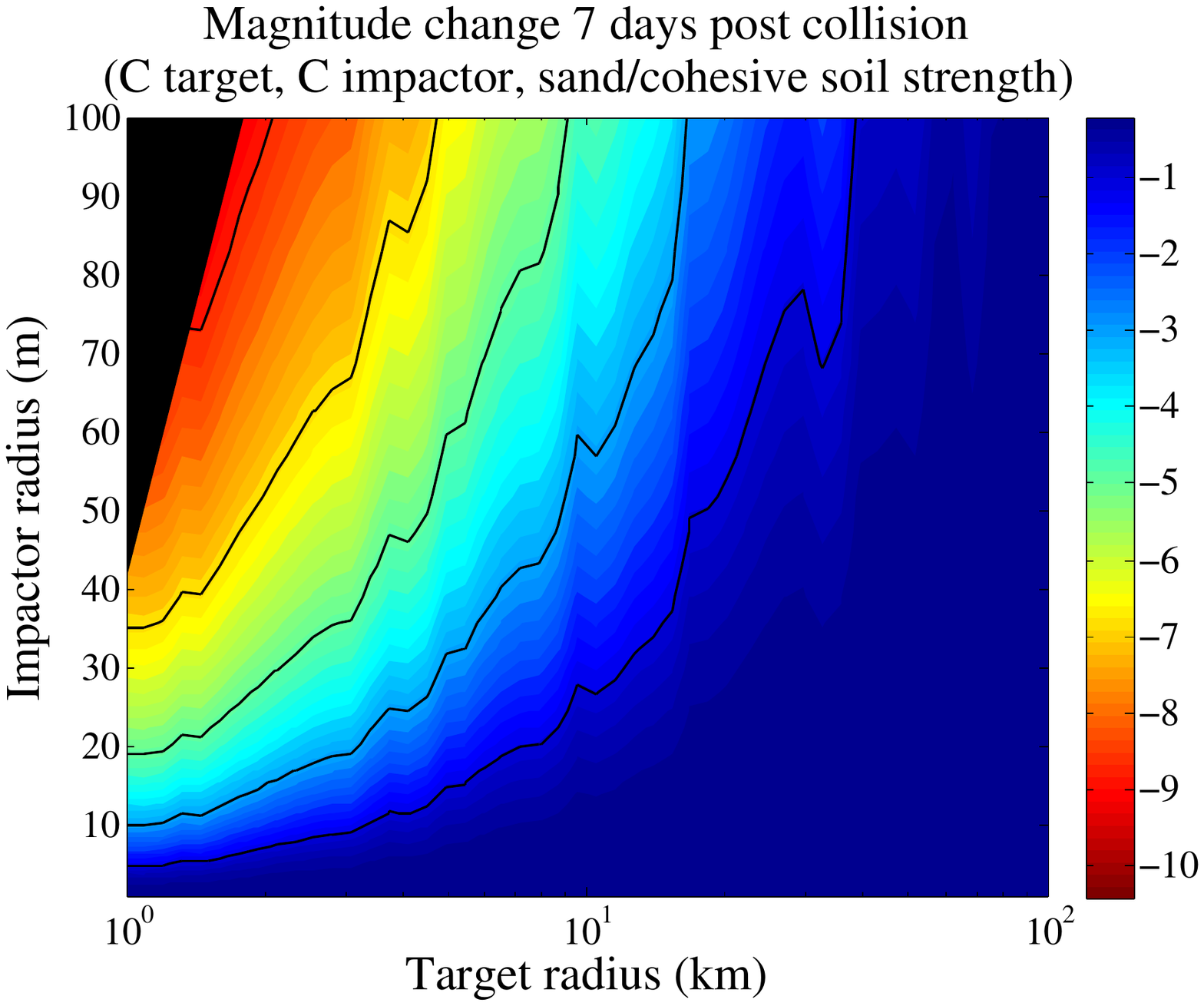}
      }

    \end{center}
    \caption{
        C-type impactor collision with C-type target: \ref{fig:CHP24}-\ref{fig:CHP1w} highly porous strength regime: magnitude change 1 and 7 days post-collision for a range of target and impactor radii. \ref{fig:CSAND24}-\ref{fig:CSAND1w} sand/cohesive soil strength regime: magnitude change 1 and 7 days post-collision for a range of target and impactor radii. Impactor radii range from 1 to 100 m, target radii: 1-100 km. Colour bar shows the magnitude change. Catastrophic disruption region is marked in black. Region where optical thickness of the debris is potentially significant is above the dashed line.  Solid black lines indicate contours where magnitude change is -1, -3, -5, -7 and -9.  }%
   \label{fig:subfiguresCHPSAND}
\end{figure}

\begin{figure}[ht!]
     \begin{center}
        \subfigure[highly porous regime, 1 day]{
            \label{fig:CHPm-124}
            \includegraphics[trim = 15mm 50mm 23mm 60mm, clip,width=0.45\textwidth]{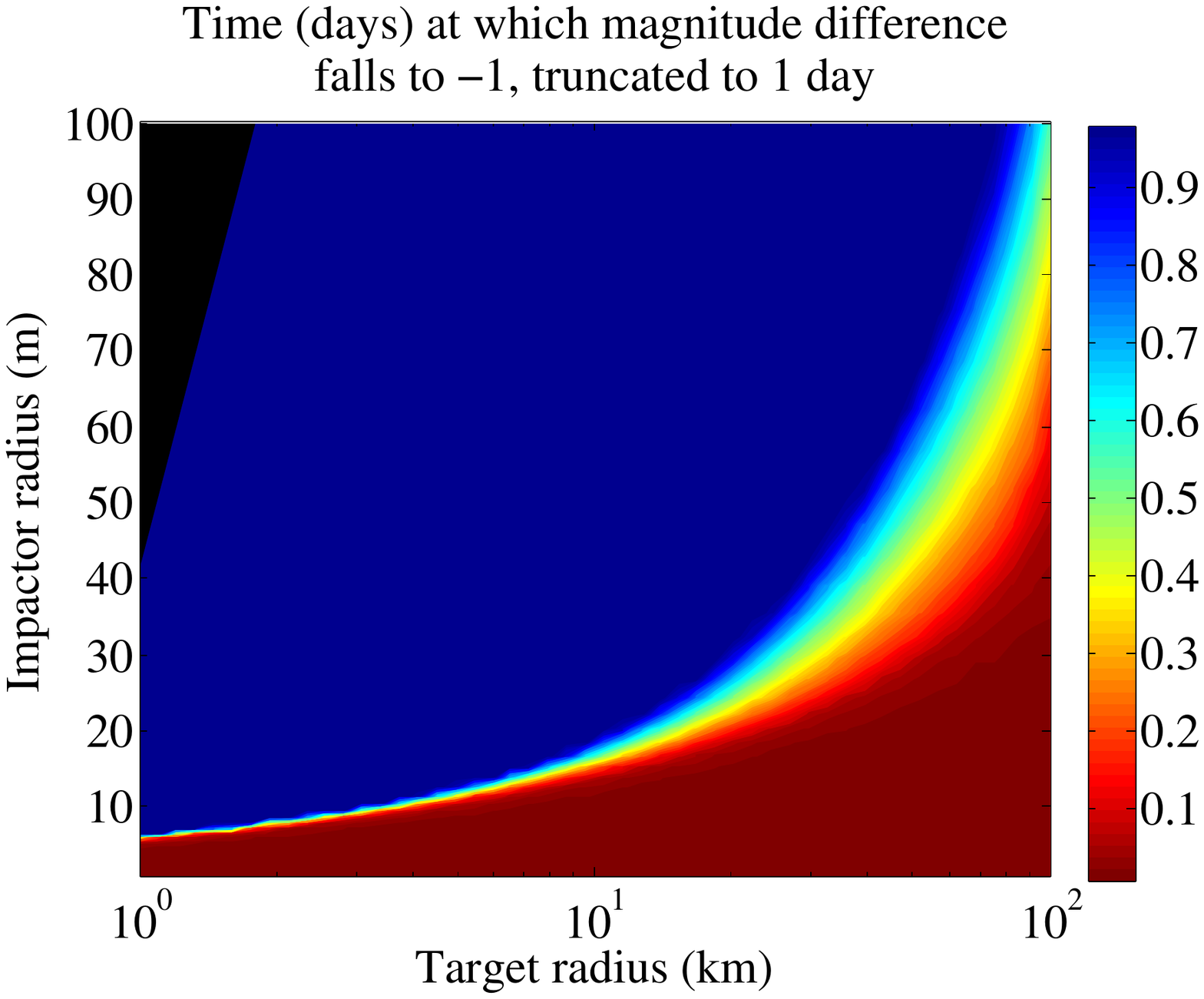}
        }
        \subfigure[highly porous regime, 7 days]{
           \label{fig:CHPm-11w}
           \includegraphics[trim = 15mm 50mm 23mm 60mm, clip,width=0.45\textwidth]{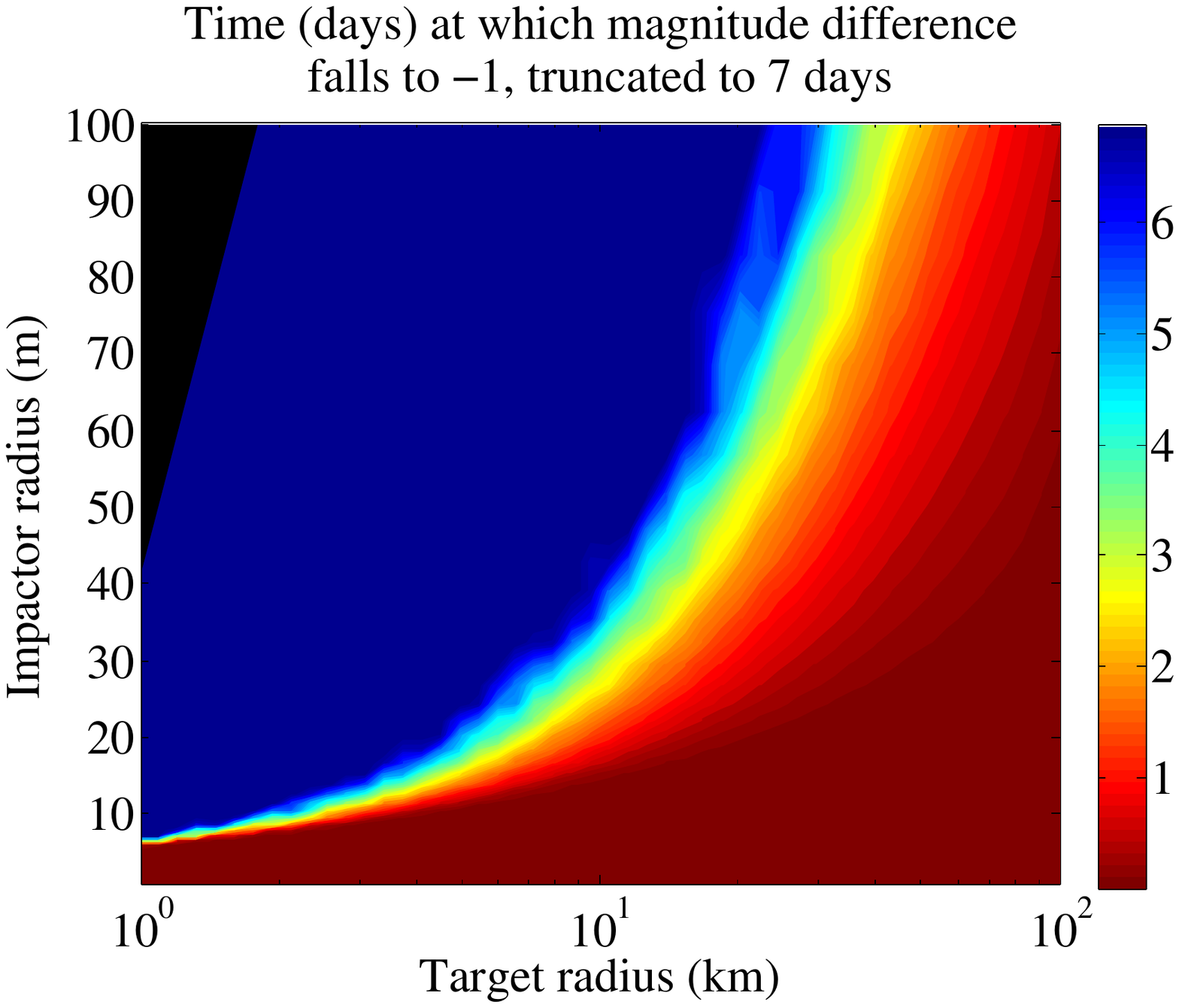}
            }
	\subfigure[sand/cohesive soil regime, 1 day]{
            \label{fig:CSANDm-124}
            \includegraphics[trim = 15mm 50mm 23mm 60mm, clip,width=0.45\textwidth]{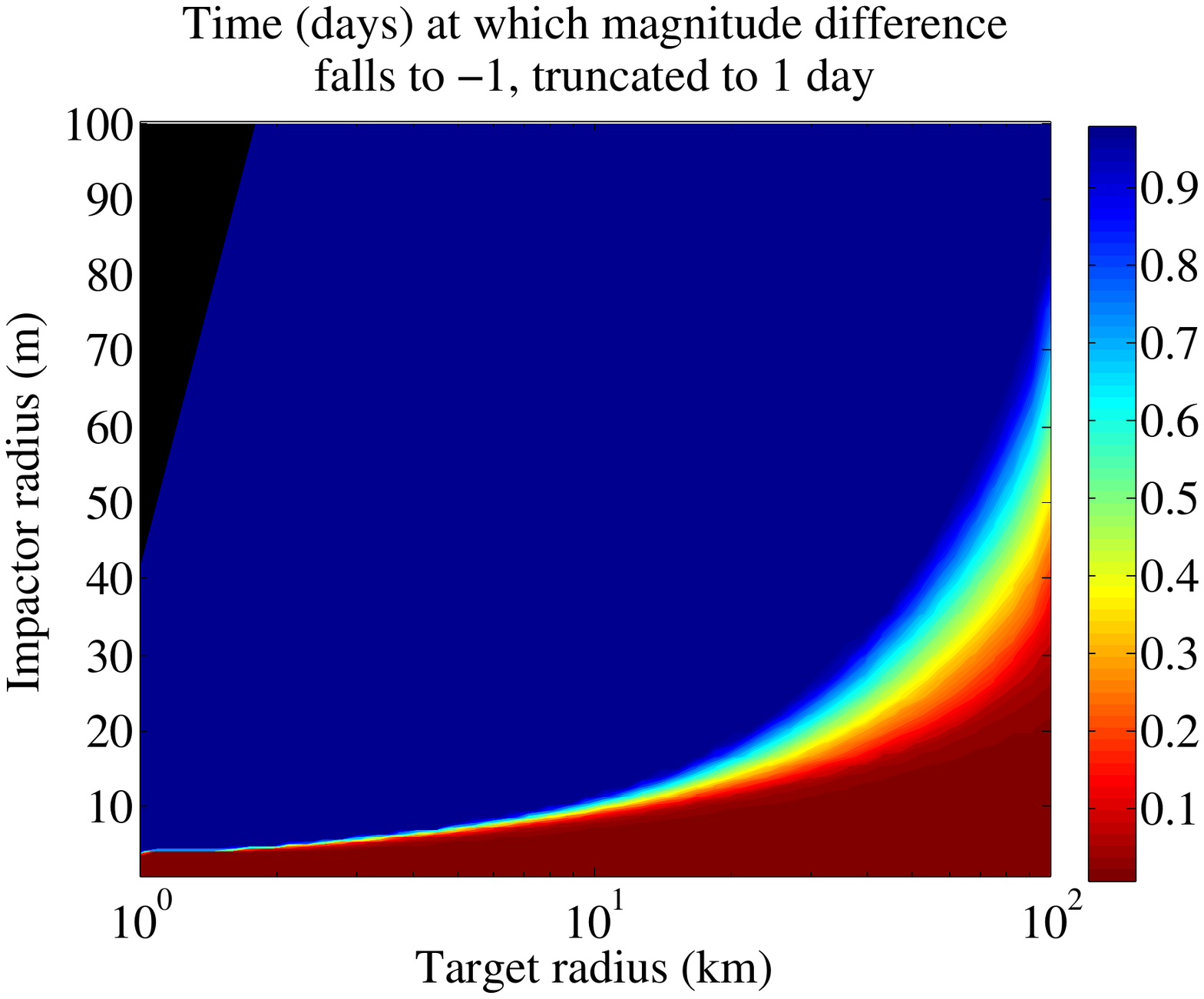}
        }
        \subfigure[sand/cohesive soil regime, 7 days]{
           \label{fig:CSANDm-11w}
           \includegraphics[trim = 15mm 50mm 23mm 60mm, clip,width=0.45\textwidth]{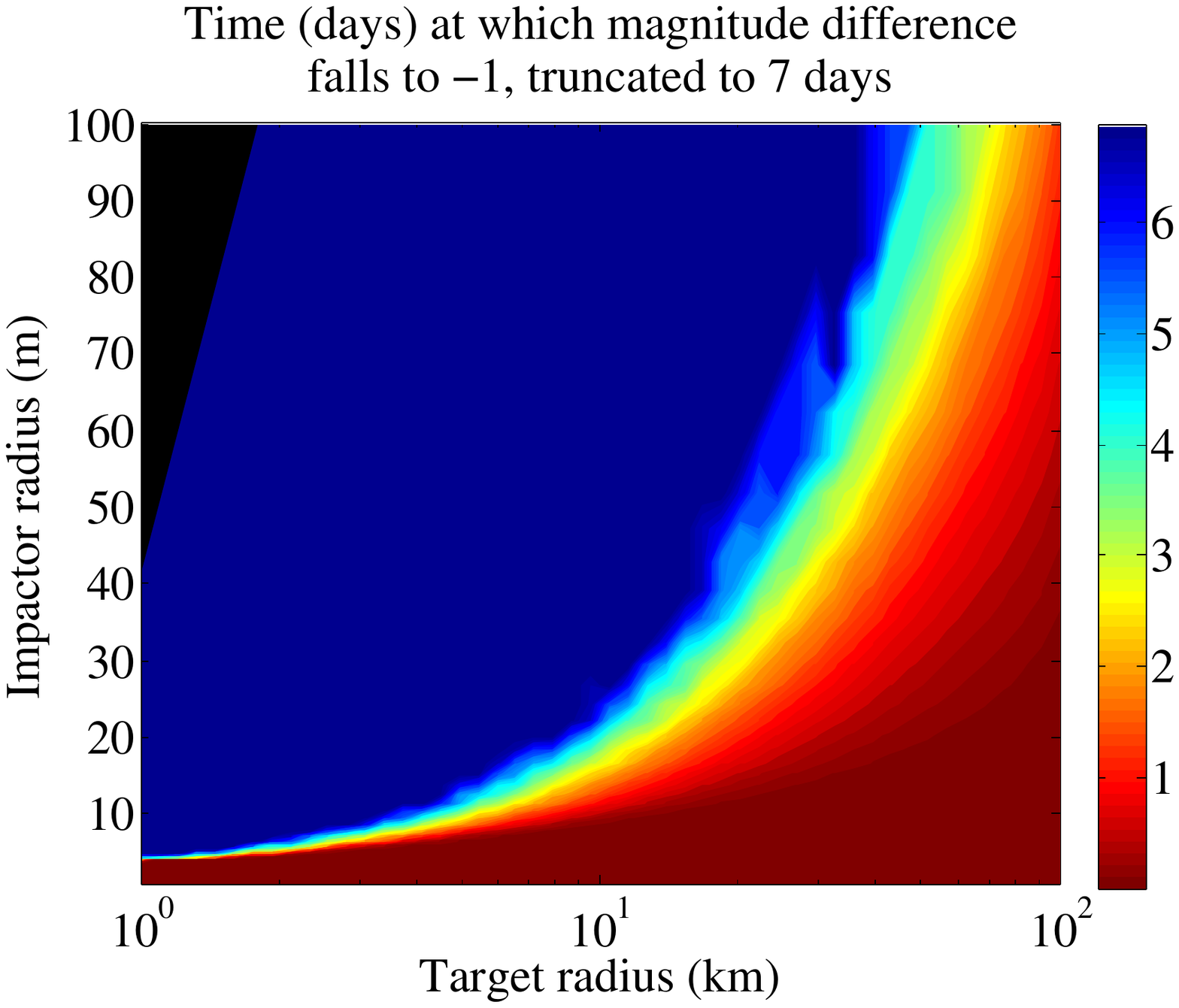}
                }

    \end{center}
    \caption{
         Time taken for the magnitude change to fall to -1 magnitudes after a C-type impactor collision with C-type target. Figures \ref{fig:CHPm-124} and \ref{fig:CSANDm-124} show the impactor-target regime where this occurs within 1 day for highly porous and sand/cohesive soil strength regimes respectively. Figures \ref{fig:CHPm-11w}-\ref{fig:CSANDm-11w} show the same data but now extended to show where this occurs within 7 days of impact. For each figure the colour coding for this timescale is shown on the right, with the truncated region (timescale greater than 1 day or 7 days) shown in dark blue on the plot. The region where a catastrophic disruption would occur is marked in black.  }%
   \label{fig:subfiguresCHPSANDm-1}
\end{figure}

Figure~\ref{fig:subfiguresCHPSAND} shows the resulting predicted change in magnitude post-collision for a C-type asteroid with both a highly porous and a sand/cohesive soil strength at two epochs after the impact (1 and 7 days). Figure~\ref{fig:subfiguresCHPSANDm-1} shows the calculated time it takes for the brightness increase to reach $-1.0$ magnitudes for C-types in the highly porous and the sand/cohesive soil strength regimes. This value has been chosen because the one sub-catastrophic asteroid collision confirmed so far between the known main-belt asteroid (596) Scheila and an impactor was observed to have a brightness increase of $\simeq-1$ magnitude~\citep{jewitt}.  Conclusive attribution of a collisional nature to a smaller magnitude decrease may be difficult, due to natural variations in brightness caused by rotational modulation of the light curves and uncertainty in asteroidal absolute magnitudes~\citep{pravecbias}. 

\begin{figure}[ht]
 \centerline{\includegraphics[trim = 12mm 50mm 23mm 60mm, clip, width=0.6\textwidth]{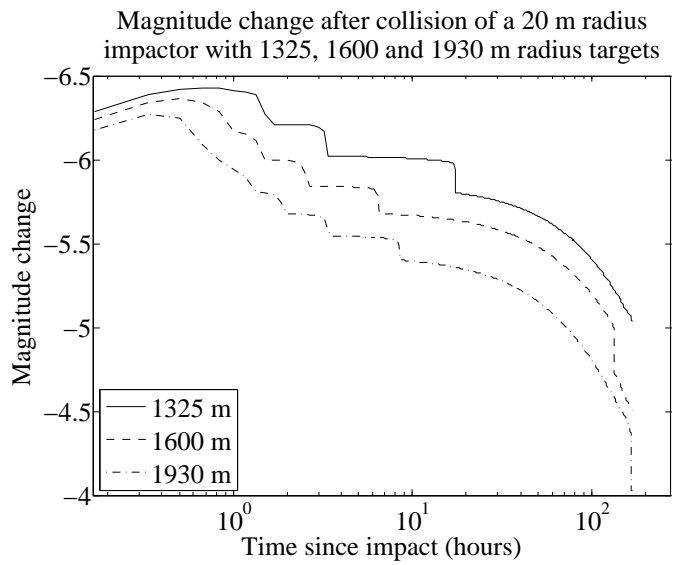}}
 \caption{Magnitude change with time for three targets of radii 1325, 1600 and 1930 m after collision with 20 m radius impactor (all S-type, sand/cohesive soil regime). Steps are clearly visible due to quantised nature of the modelled velocity shells and occur at a different position for each target. }
 \label{fig:steps}
\end{figure}
\clearpage

There are vertical structures in these plots that are not of physical significance, but are rather features of the model limitations. Three processes affect the total magnitude decrease post collision: the constant expansion of debris moving out of the aperture, material falling back on the surface of the target asteroid and therefore disappearing from observation; and material re-entering the aperture due to the gravitational attraction of the target asteroid. In an ideal simulation of the process, the change in magnitude as a function of time would be a smooth function, being the result of the interplay of all three processes. The function would be smooth because we expect the size and velocity distributions of the debris to be continuous. As velocity shells in this model are quantised, this leads to the creation of artefacts at certain values of target radius.

The effect of this can be seen clearly in Figure~\ref{fig:steps}, which shows the predicted brightness increase after an impact of a 20 m radius impactor onto 1325, 1600, 1930 m radius target (all S-type, sand/cohesive soil strength regime). For lower velocity shells that fall back onto the asteroid at early times these steps merge to give a continuous slope. Higher velocity shells never return as they have reached escape velocity, and thus never produce a sharp decline. 

For a small number of shells with intermediate velocity, the return of the shell to the surface will be sufficiently distinct temporally from the other shells to produce a clearly identifiable step.
The velocity shells leaving the aperture do not produce a sharp decline, because the fraction of the shell in the aperture decreases as a continuous function of time after the impact.
The location of these steps is independent of the impactor radius and depends solely on the target radius due to the dependance on the surface gravity. These steps are clearly visible in Figures \ref{fig:subfiguresCHPSAND} and after as vertical structures that should not be interpreted as being significant. 

\begin{figure}[ht]
     \begin{center}

        \subfigure[7 days, 500 shells]{%
            \label{fig:CCsand500_7d}
            \includegraphics[trim = {\myleft} {\mybottom} {\myright} {\mytop}, clip,width=0.45\textwidth]{C_s_C_7days_new_density_500.pdf}
        }
        \subfigure[7 days, 5000 shells]{
            \label{fig:CCsand5000_7d}
            \includegraphics[trim = {\myleft} {\mybottom} {\myright} {\mytop}, clip,width=0.45\textwidth]{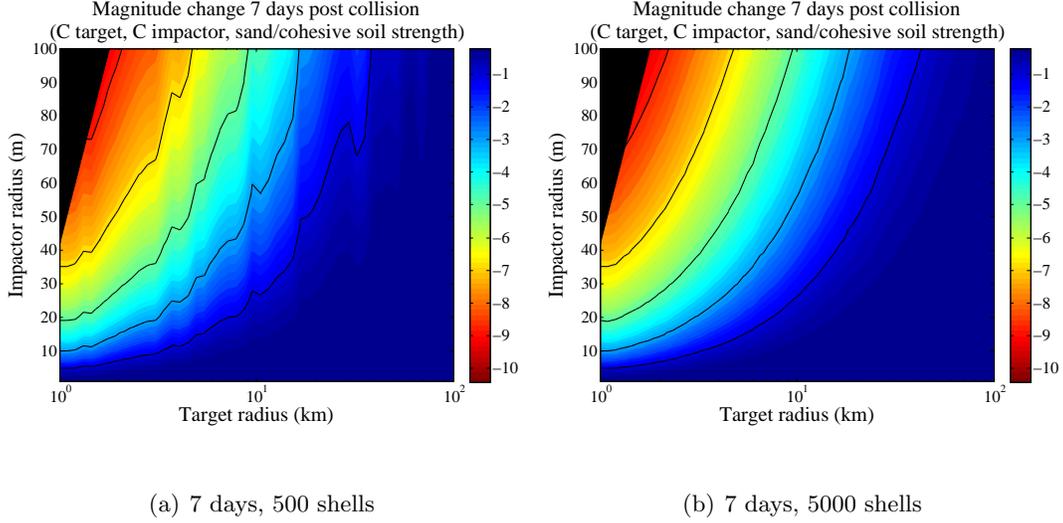}
            }

    \end{center}
    \caption{
          C-type impactor collision with C-type target, sand/cohesive soil strength regime: magnitude change 7 days post-collision for a range of target and impactor radii. Impactor radii range from 1 to 100 m, target radii: 1-100 km. Colour bar shows the magnitude change. Catastrophic disruption region is marked in black. Comparison of using (a) 500 and (b) 5000 velocity shells. The plot with 5000 shells shows a clear reduction in the vertical structures in comparison to 500 shells plot. Solid black lines indicate contours where magnitude change is -1, -3, -5, -7 and -9.}%
   \label{fig:subfiguresCCsand_shells_comparison}
\end{figure}

One of the ways to minimise this and have a good resolution in the images is by using a large number of shells, however this increases the computation time. The 500 shells used in this study was selected as a compromise between resolution and run time. Figure \ref{fig:subfiguresCCsand_shells_comparison} shows a comparison between a plot using 500 shells and 5000 shells (20 hours run-time). The latter shows a reduction in vertical structures without substantial difference in output. Difference images show a discrepancy between the 500 shells and 5000 shells results being confined to the vertical structures only.

We have chosen to only run the model up to 7 days post-collision, because a large proportion of the material (by surface area) would leave the aperture due to radiation pressure effects, which are not currently included in the model.
The time  $t$ it takes for material to escape the aperture radius $r_{ap}$ is approximately equal to
\begin{equation}
t\approx\left(\frac{2r_{ap}}{a}\right)^{\frac{1}{2}}
\end{equation}

where $a$ is acceleration, that can be defined in terms of ratio between radiation pressure acceleration and the local gravity, $\beta$,  and gravitation acceleration to the Sun at 1 AU $g_{\odot}$: $a=\beta g_{\odot}$. 
In turn, $g_{\odot}\approx \frac{GM_{\odot}}{{R_{h}}^2}$, where $R_{h}$ is heliocentric distance.

This gives the approximate time for radiation pressure to remove material from the aperture as

\begin{equation}
t\approx\left({\frac{2r_{ap} {R_{h}}^2}{\beta GM_{\odot}}}\right)^\frac{1}{2}
\end{equation} 

where $R_{h}$ is distance from the sun to the asteroid in question.

Assuming $\beta$ is on the order of $\approx 0.1$\citep{beta} for the small grains in the ejecta expected to dominate the scattering, $R_{h}=3$ AU (mid-main belt), the time $t$ is approximately 2 days. Therefore, we do not present any model results that go beyond 7 days post collision, as without the inclusion of the radiation pressure they do not reflect physical reality.

\subsection{Optical depth effects}

Throughout the modelling, all ejecta within the aperture and not behind the asteroid is treated as contributing to the brightness increase via the projected surface area. Effectively, it is assumed that optical depth $\tau$ is sufficiently low that the approximation $\tau\simeq1-e^{-\tau}$ is valid. (For $\tau <0.2$ this leads to a brightening overestimate of less than 10\%). 
To estimate the potential optical depth effects, the ejecta cloud was divided into a series of concentric rings and the column density of each was calculated, then converted, using the particle size distribution, into the optical depth $\tau$. Any rings with $\tau < 0.02$ were ignored (this corresponds to an error of less than 1\%), and any cases where more than 10\% of ejecta mass was contained in rings with $\tau>0.02$ were considered to have their brightness potentially overestimated. We found that at one day after impact, ignoring opacity may lead to overestimating the brightness increase by up to twenty percent ($\sim 0.2$) magnitudes.  The region where this may be relevant is outlined in shown in figures  \ref{fig:CHP24}, \ref{fig:CSAND24}, \ref{fig:SSAND24} and \ref{fig:DSAND24} showing the brightness 1 day post collision. Clearly for all impacts apart from those on the largest asteroids, the correction to the calculated magnitude change will have little effect on detection and can be ignored. By one week after impact, we calculate that optical depth effects are negligible for all target impactor pairs.

\section{Discussion}

A collision between asteroids is a multi parameter problem requiring us to make some initial assumptions. This section is therefore divided into 4 parts. 
First we describe an application of the model to the only known collision in the  parameter space explored - (596) Scheila. In the second part we examine the effect of the different strength regimes in a collision between C-type asteroids (as the most numerous in the outer belt where (596) Scheila is situated). In the third part we keep the strength regime the same (sand/cohesive soil) and vary the taxonomic types to examine the effect it has on the magnitude change. We conclude with application of the model to predicting the type of observable collisions by currently active surveys such as Pan-STARRS 1 and the Catalina Sky Survey.

\subsection{Modelling the magnitude decrease in the (596) Scheila collision}

In December 2010 asteroid (596) Scheila was observed with the 0.68 meter Catalina Schmidt telescope to have increased in brightness by approximately 1.3 magnitudes in comparison to the previous month \citep{larson}.  There have been multiple photometric observations of (596) Scheila reported as summarised in Table~\ref{tab:scheila_obs}, from which the impact is estimated to have occurred on December 3.5 UT, 2010 \citep{ishiguro_obs}.
All observations measured the total magnitude, i.e. from the asteroid and the surrounding debris. \cite{jewitt} used a 105,600 km aperture for the HST observations of (596) Scheila and this value is used for the calculations throughout this section. We have selected these observations for comparison with our model as giving the true magnitude decrease because they were taken by the same instrument and in the same observational circumstances. 

\begin{table}[ht] 
\caption{Summary of observed change in magnitude for (596) Scheila} 
\centering 
\begin{tabular}{c c c } 
\hline\hline 
Date of observation (UT) & Approximate time & Magnitude  \\  
& post-collision (days) & change  \\[0.5ex]
\hline 
2010 Dec. 3.4 \citep{larson} & 0& -1.3 \\ 
2010 Dec. 11.44-11.47 \citep{larson}& 7.9 & -1.1 \\ 
2010 Dec. 12 \citep{hsieh}& 9 & -0.86 \\ 
2010 Dec. 14-15 \citep{bodewits2}& 11 & -0.66 \\ 
2010 Dec. 27.9 \citep{jewitt} & 24.4 & -1.26 \\ 
2011 Jan. 4.9 \citep{jewitt} & 32.4 & -1.00 \\ [1ex] 
\hline 
\end{tabular} 
\label{tab:scheila_obs} 
\end{table}

A collision with a smaller asteroid is the most likely cause of this brightening and there have been a range of impactor diameters suggested in literature: 35 m \citep{jewitt}, 30-50 m \citep{ishiguro}, 30-90 m \citep{moreno}, $<$100 m \citep{bodewits}. 
The impactor diameters above are calculated using the estimated mass of high speed ejecta and the relationship $m(v) \approx m_{i}$ for $v = 10^{-2} v_{i}$ \citep{HH2011} where $m(v)$ is the mass of ejecta above velocity $v$, $m_{i}$ is the impactor mass and $v_{i}$ the impactor velocity.

\begin{figure}[ht!]
     \begin{center}
        \subfigure[C-type impactors]{
            \label{fig:Cimp}
            \includegraphics[trim = 18mm 50mm 25mm 60mm, clip,width=0.45\textwidth]{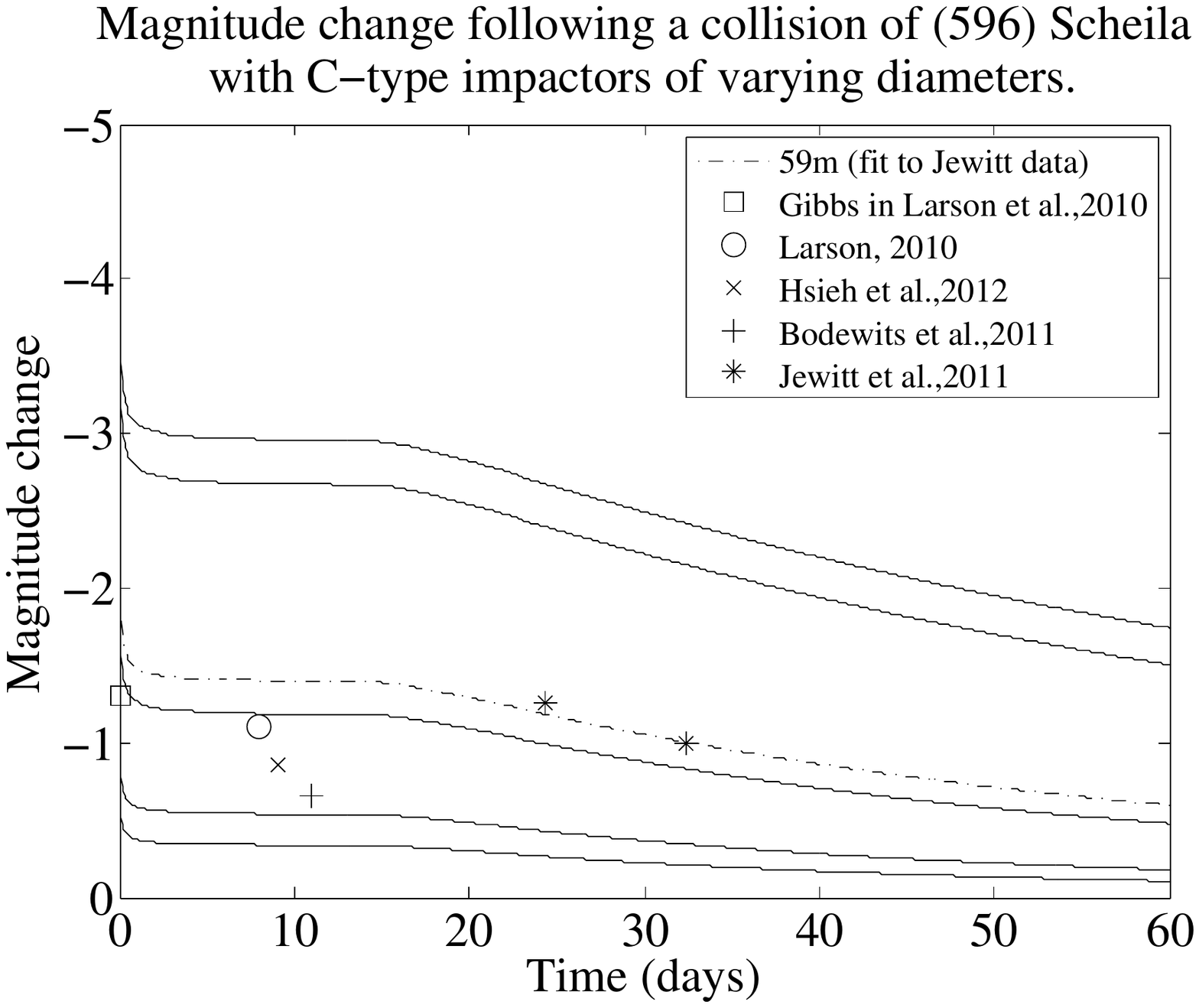}
        }
        \subfigure[S-type impactors]{
           \label{fig:Simp}
           \includegraphics[trim = 18mm 50mm 25mm 60mm, clip,width=0.45\textwidth]{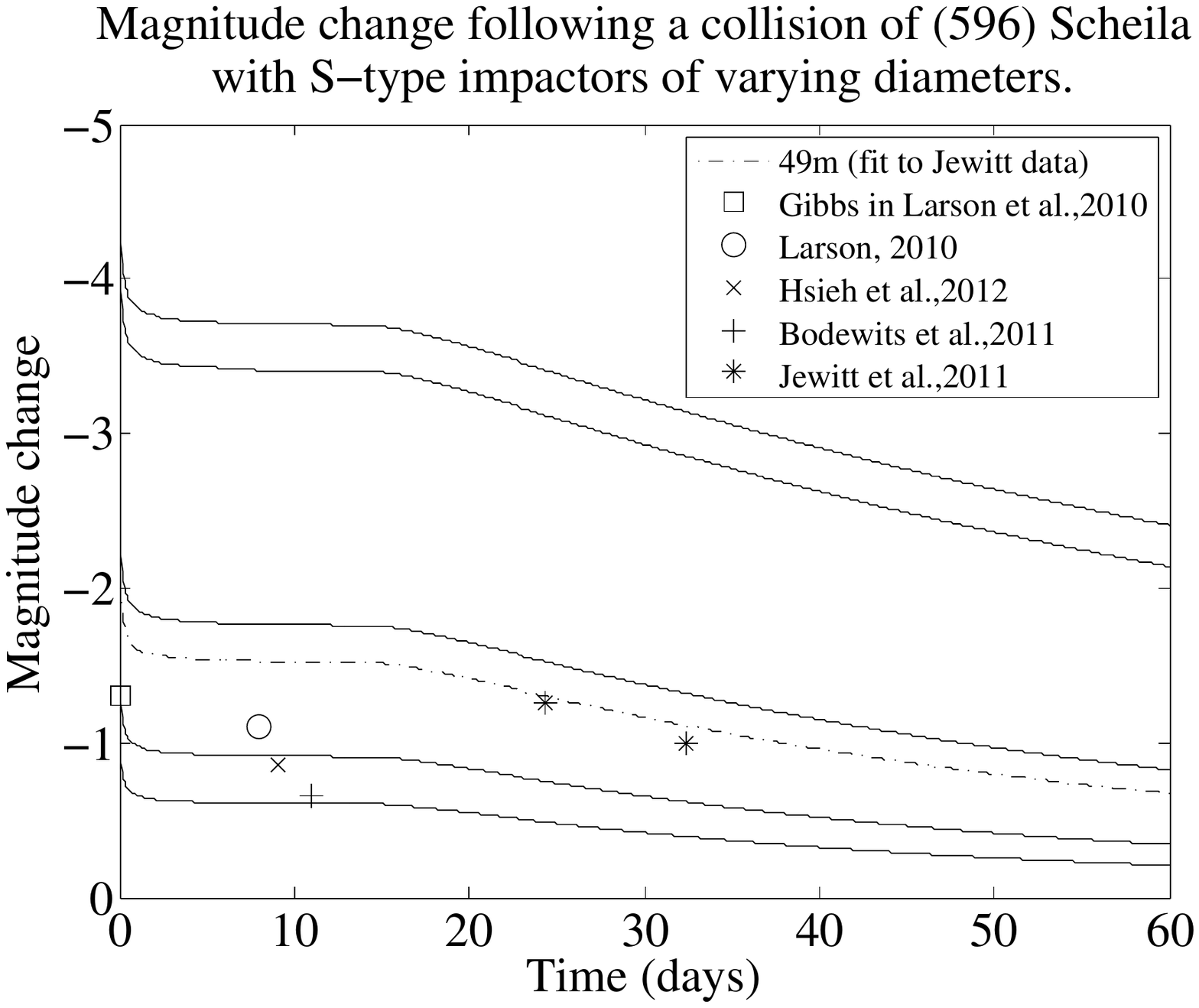}
        }\\ 
        \subfigure[D-type impactors]{%
            \label{fig:Dimp}
            \includegraphics[trim = 18mm 50mm 25mm 60mm, clip,width=0.45\textwidth]{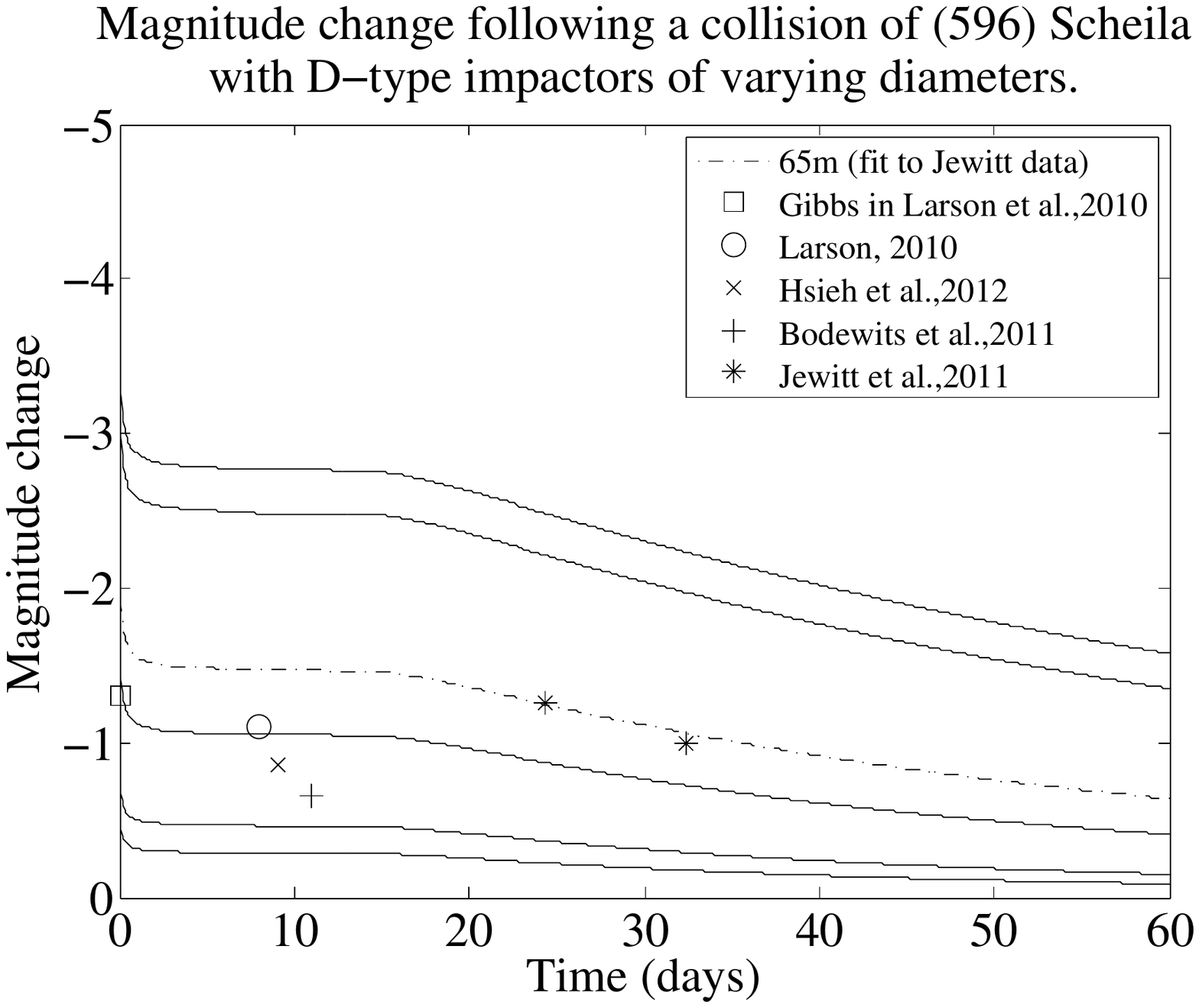}

        }

    \end{center}
    \caption{
        Magnitude change as a function of time following a collision of C-, S- and D-type impactors with a range of diameters with asteroid (596) Scheila. Individual points show reported magnitude change by various authors. Dot-dash line shows our fit to Jewitt et al.'s data. The best fit estimates of impactor diameter with our model are 59, 49 and 65m for C-, S- and D-type impactors respectively.  }%
   \label{fig:subfigures_scheila}
\end{figure}

Scheila has semimajor axis a = $2.926$ AU, eccentricity $e = 0.164$, and inclination $i = 14.7^{\circ}$, which places it in the outer main belt. The most numerous asteroids in that region and therefore the most likely impactors are C-type \citep{taxtypes}, although (596) Scheila itself is a D-type \citep{dtype}. The taxonomic type determines the densities used in the model which affect the final output parameter of change in magnitude. 
Figure ~\ref{fig:subfigures_scheila} shows the post-collision magnitude change during 60 days for a collision between Scheila and C-, D- and S-type impactors of diameters reported in the literature. We also used the observations from \cite{jewitt} to find the best fit size for each impactor type by matching the calculated magnitude change in our model to the observed lightcurve. The results are 59 m, 49 m and 65 m diameter impactors for S-, C- and D-types respectively. 
As has been mentioned in the previous section, we believe that in its current state the model gives best description of reality if we limit the results to 7 days post-collision.  However, in the case of Scheila collision, which is an important real observed collisional event, we lack a sufficient number of data points of good quality in the first 7 days to make a meaningful comparison. However, the large aperture used in this study includes a large portion of the ejecta even after being significantly affected by radiation pressure, allowing comparison with our model.
The model results are in approximate agreement with estimated impactor diameters by other authors. The differences most likely come from a variety of initial assumptions made by the different researchers, i.e. density, type of ejecta particle distribution, range of particle sizes and the estimated ejected mass. It is also clear from these results, that to the first order the taxonomic type of the impactor asteroid is not an important factor in diameter estimation.

\subsection{Effect of strength regime on post-collision magnitude change for C-type asteroids.}

We consider a collision where both impactor and the target are generic C-type asteroids. Figures~\ref{fig:CHP24} and \ref{fig:CHP1w} show the magnitude decrease post-collision in highly porous regime, while Figures~\ref{fig:CSAND24} and \ref{fig:CSAND1w} show the effect of changing the regime to the sand/cohesive soil.
Following the collision, a shock wave travels through the target asteroid and the outcome of the collision  depends on how the stress wave propagates and is attenuated through the target. The cratering and amount of ejecta following an impact is determined by the target internal structure, porosity and strength. A porous asteroid would react to the impact differently than a more solid body and it is particularly important to consider impacts on these  as most observable
main belt asteroids should have significant macroporosities, apart from the very largest. It has been found that highly porous objects such as (253) Mathilde are particularly good at attenuating stress waves due to energy going into the compaction of the pores \citep{porosity}.
 An object with large macroporosity is most likely to be highly fractured or even a rubble pile. Dark and primitive asteroids are more likely to have significant porosity, however it is worth noting that it is not exclusive to C-types.  
Figures~\ref{fig:CHP24}/\ref{fig:CHP1w} and \ref{fig:CSAND24}/\ref{fig:CSAND1w} are similar for both strength regimes at each point of time considered, however the sand/cohesive soil regime shows a larger magnitude decrease due to more material ejected, while in the highly porous regime some of the impact energy goes into compaction.
This tendency is also reflected in plots of amount of time taken for the magnitude change to decrease to $1.0$ magnitudes above the pre-impact brightness (Figure \ref{fig:subfiguresCHPSANDm-1}). The brightness increase caused by the ejecta reaches the assumed observable limit of approximately 1 magnitude faster for the highly porous case than the corresponding impactor-target pair undergoing a collision in the sand/cohesive soil regime. This is due to less ejecta being produced.

Recent research by \cite{carry} indicates that the density may be dependant on target size, rather than being a fixed quantity for each taxonomic type. However, the data is sparse in the size range used in this model and unavailable for the D-types. Therefore, our model makes a simplifying assumption that the bulk density stays fixed for all sizes considered and depends only on the taxonomic type.

\subsection{Effect of taxonomic type (C, D and S) on post-collision magnitude change in sand/cohesive soil regime}
\begin{figure}[ht!]
     \begin{center}
        \subfigure[1 day]{
            \label{fig:SSAND24}
            \includegraphics[trim = {\myleft} {\mybottom} {\myright} {\mytop}, clip, width=0.45\textwidth]{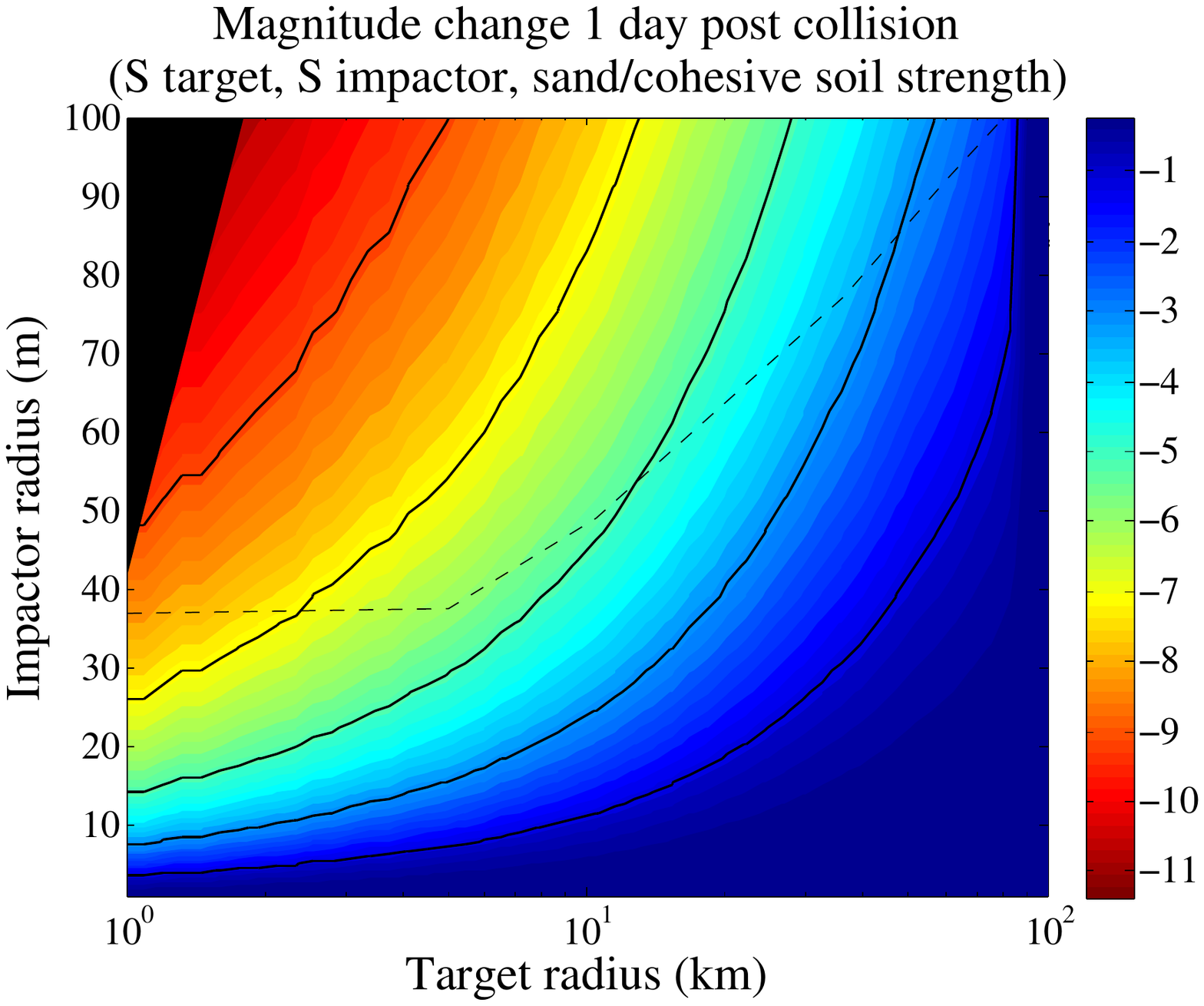}
        }
        \subfigure[ 7 days]{
           \label{fig:SSAND1w}
           \includegraphics[trim = {\myleft} {\mybottom} {\myright} {\mytop}, clip,width=0.45\textwidth]{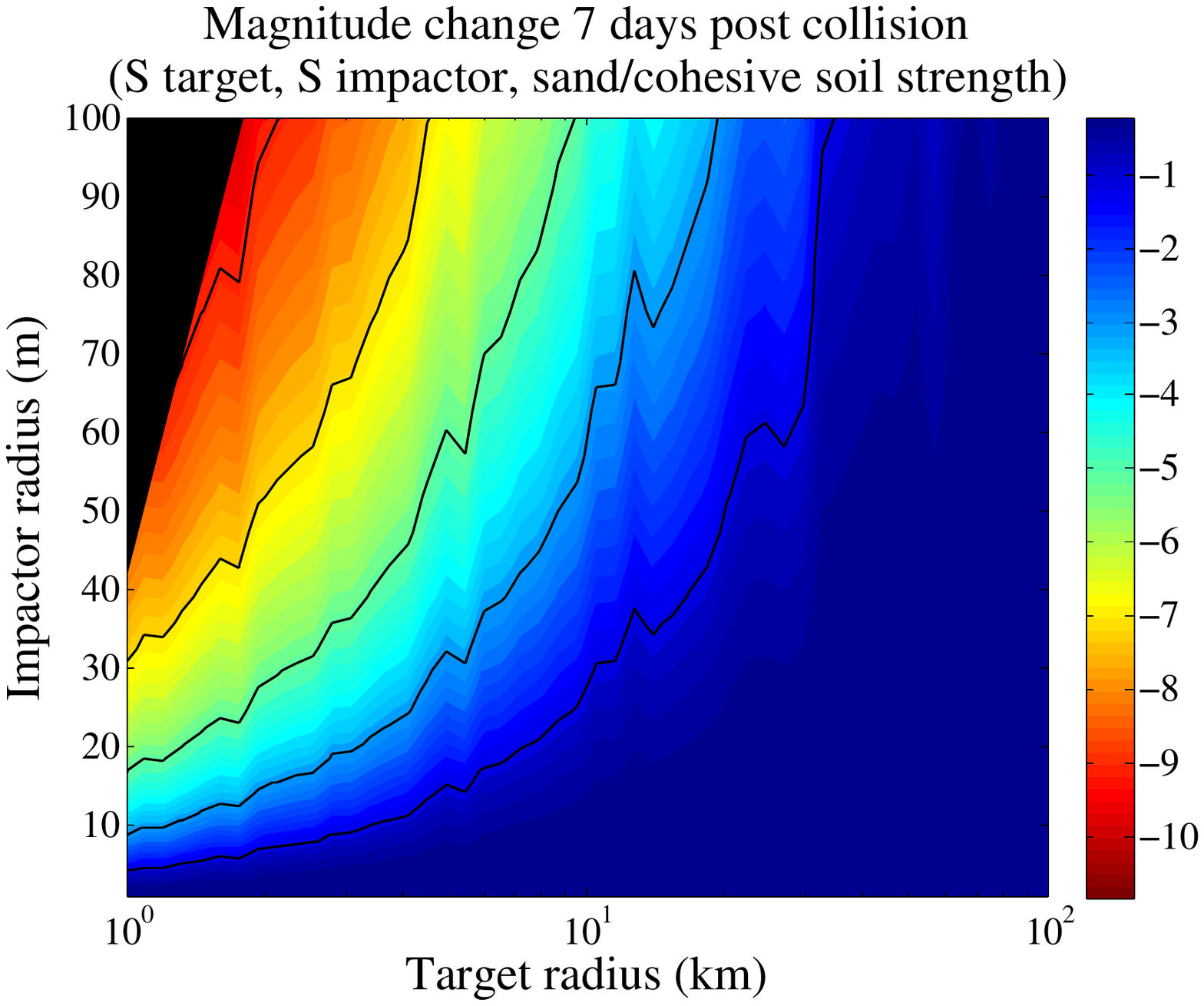}
               }

    \end{center}
    \caption{
         S-type impactor collision with S-type target, sand/cohesive soil strength regime: magnitude change 1 and 7 days post-collision for a range of target and impactor radii. Catastrophic disruption region is marked in black. Region where optical thickness of the debris is potentially significant is above the dashed line. Solid black lines indicate contours where magnitude change is -1, -3, -5, -7 and -9. }%
   \label{fig:subfiguresSSAND}
\end{figure}

\begin{figure}[ht!]
     \begin{center}
        \subfigure[1 day]{
            \label{fig:DSAND24}
            \includegraphics[trim = {\myleft} {\mybottom} {\myright} {\mytop}, clip, width=0.45\textwidth]{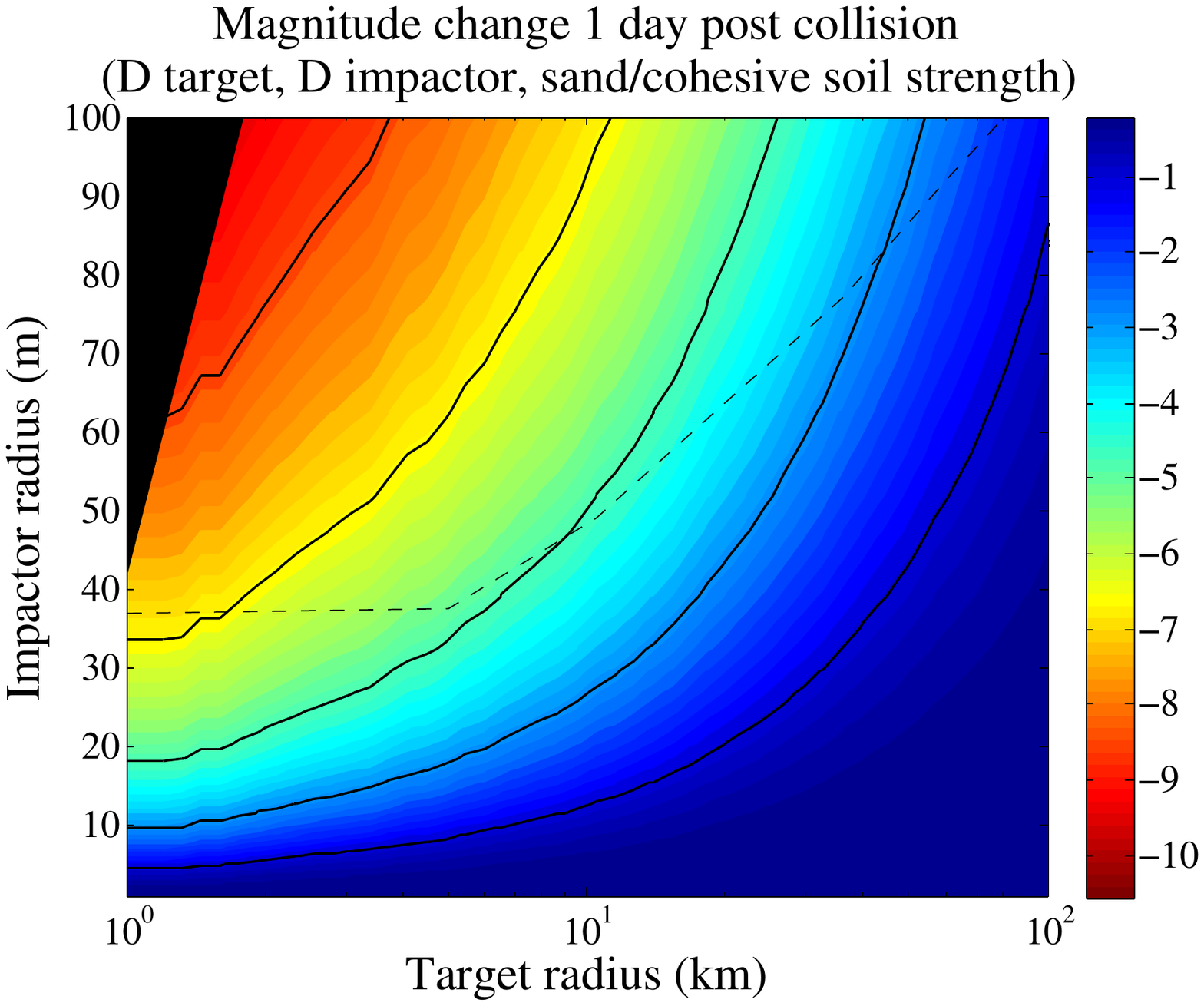}
        }
        \subfigure[ 7 days]{
           \label{fig:DSAND1w}
           \includegraphics[trim = {\myleft} {\mybottom} {\myright} {\mytop}, clip,width=0.45\textwidth]{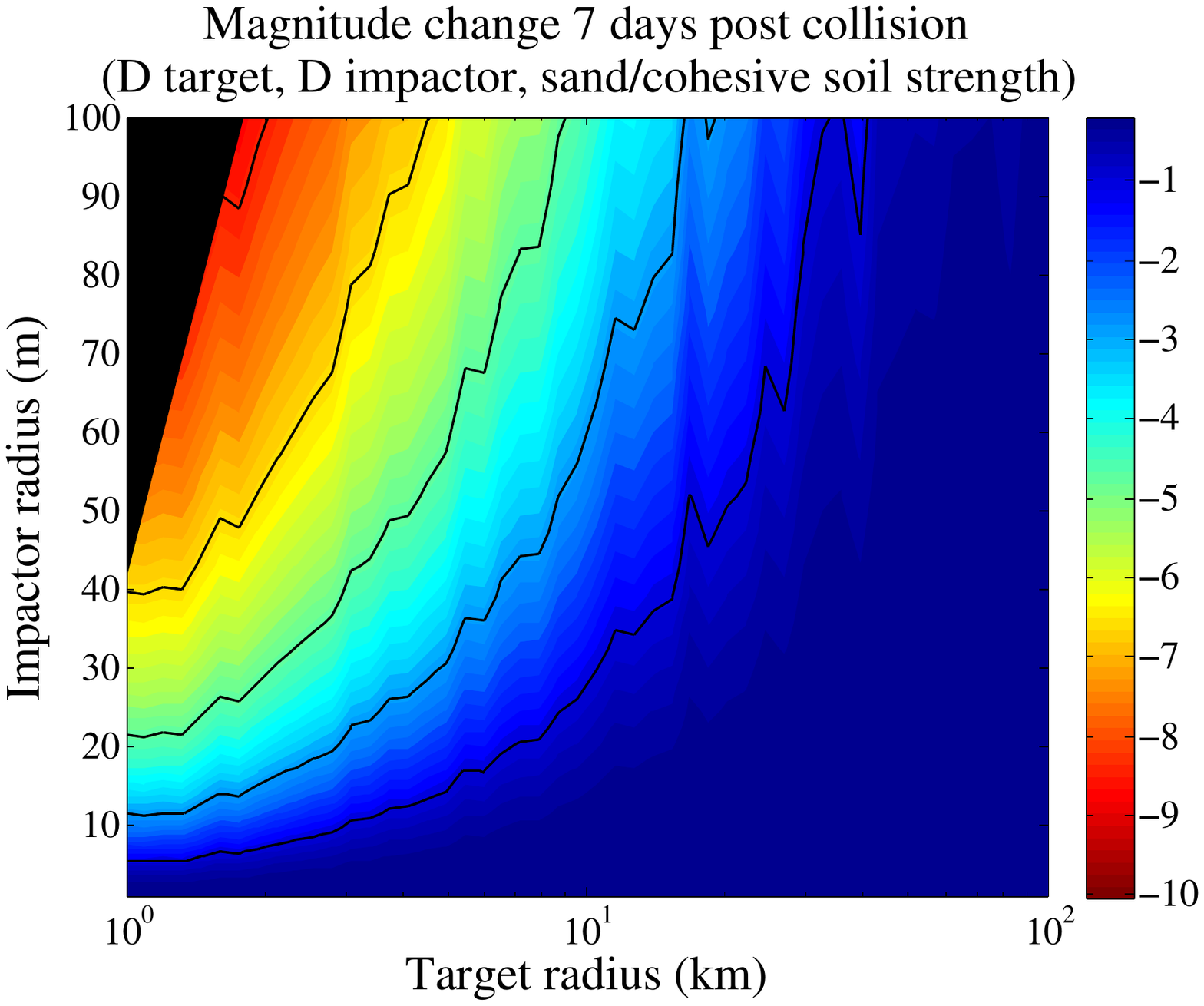}
             }

    \end{center}
    \caption{
         D-type impactor collision with D-type target, Sand/cohesive soil strength regime: magnitude change 1 and 7 days post-collision for a range of target and impactor radii. Catastrophic disruption region is marked in black. Region where optical thickness of the debris is potentially significant is above the dashed line. Solid black lines indicate contours where magnitude change is -1, -3, -5, -7 and -9.  }%
   \label{fig:subfiguresDSAND}
\end{figure}

It is important to understand the taxonomy effects on the collision lightcurve due to the established composition gradient across the main belt.
To investigate we kept the strength regime (sand/cohesive soil) constant and varied taxonomic type of both target and impactor (for S-, C- and D-types). Figures \ref{fig:subfiguresSSAND} and 
\ref{fig:subfiguresDSAND} show the magnitude change 1 and 7 days post-collision for target-impactor pairs. Other combinations of target-impactor types are included in the Supplementary material (available online).
S-types impacting each other are more likely to occur in the inner belt, while D-types impacting each other are more likely in the Trojan population. Generally, both sets of plots show qualitatively similar results for magnitude change with time, with a high magnitude change in the day 1, followed by the decline over time, with C-types having the least magnitude increase at 7 days. From our results we observed that collisions between the lowest grain density and highest porosity C-types produce the least amount of magnitude decrease lasting for less time that it does for other taxonomic types. The S-types which have lower porosity and the highest considered grain density show the opposite - a larger magnitude change lasting longer. D-type collisions fall between these regimes. 

\begin{figure}[ht]
\centerline{\includegraphics[trim = 18mm 50mm 18mm 60mm, clip, width=0.7\textwidth]{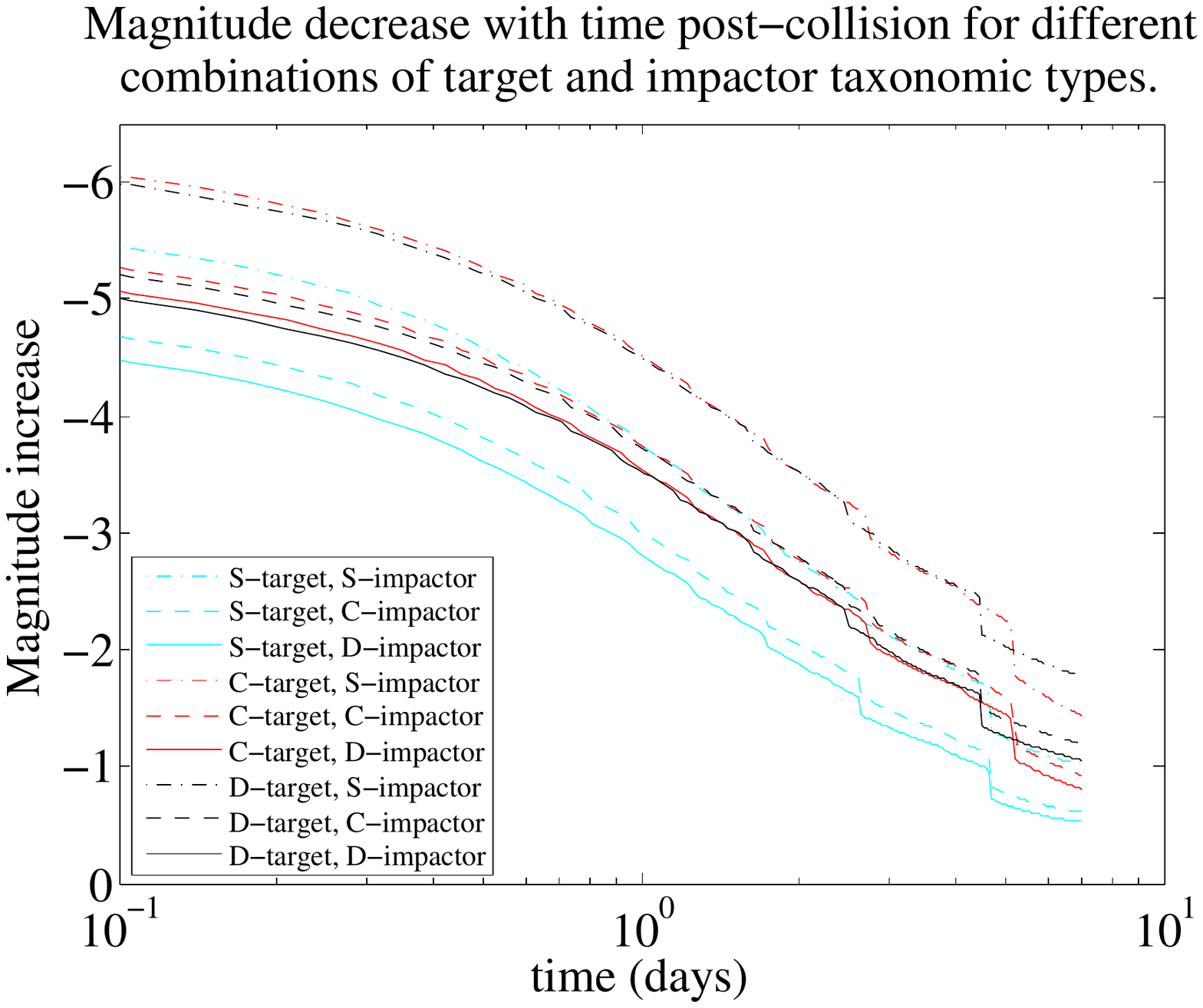}}
 \caption{The effect of varying taxonomic type in a collision of 20km target and 50m impactor on magnitude change as a function of time.}
 \label{fig:taxtypes}
\end{figure}

Figure \ref{fig:taxtypes} shows a direct comparison between different combinations of target and impactor taxonomic types for a 20 km target and 50 m diameter impactor. 
In a collision with an S-type target, the largest magnitude change is produced by a S-type impactor, then C-type and D-type producing the least magnitude change. The same pattern is followed by a C-type target being impacted by S-, C- and D-type asteroids. However, a D-type target deviates from this pattern, with an S-type impactor giving the largest change in magnitude, followed by D-types and C-types giving the least magnitude change. Overall, S-type impactors produce the brightest collision signature in all types of targets considered, due to their inherent high grain density.

\subsection{Detection from current sky surveys}
 
The  catastrophic collision rate in
the main belt has been calculated from observationally constrained dynamical models as $\sim1$ per year at a diameter $D\sim 100$ m
~\citep{durda, greenberg, Bottke2005}. The general hope within the scientific community has been that
such collisions would be detected via the on-going wide-field NEO surveys.
However, several objects initially identified as potential collisional disruption events are now
suspected of being caused by rotational disruption due to YORP spinup {\it e.g.}
P/2010 A2 ~\citep{Agarwal2013}, P/2013 P5 ~\citep{Jewitt2013}. Additionally, a
recent study by ~\cite{Denneau2014} using $1.2$ years of Pan-STARRS 1 data found only one
plausible candidate of a collision event, and concluded that collisional
disruptions of  100-m scale asteroids  may be extremely rare. Hence observing
much more frequent sub-catastrophic collisions may be a viable method for
constraining the overall collision rate.

To look at the likelihood of detecting these events, we first look at collisions with
100 km diameter targets similar to Scheila. According to \cite{Bottke2005}, the size of impactor that would disrupt a $D=100$ km asteroid
is $D\geq25$km and such a disruption would happen every $\sim10^{7}$ years in the main belt.
Using  their CoDDEM model size distribution, a subcatastrophic collision impact by a $D\geq50$ m 
impactor should occur once per $\simeq 25$ years, in rough agreement with the single such collision observed
since the start of modern surveys in the 1990's.
The current Pan-STARRS and Catalina surveys are complete down to an apparent magnitude $V\simeq 20$ (equivalent to  $D\sim 2$ km in the main-belt), but are sensitive to asteroids down to $V\simeq 22$. The Bottke et al. disruption frequency at $D\sim 2$km is approximately one per 1000 years, requiring an impactor $D\simeq 60$ m.  Scaling by the CoDDEM model size distribution predicts a collision between a $D\sim 2$km asteroid and a $D\geq10$m  impactor should occur approximately once per year. As $\sim 50$\% of the asteroid belt is visible at any time,  current surveys  should be able to detect a sub-catastrophic impact once every couple of years.

So why are these collisions between smaller bodies not found? In reality, observational surveys are affected by 
  factors such as weather, moonlight and occasional losses due to detector gaps in the cameras. Perhaps most
importantly, current surveys have a cadence for most individual asteroids measured in weeks rather than days.  Using a detection aperture of $\sim 1000$~km equivalent to $\sim 1-3$ arcsec as appropriate to surveys such as Pan-STARRS, our model predicts a brightening of $\geq-1$ magnitudes for $D\simeq10$m impacts on $D\sim 2$km for only 10--20 days. In reality, effects  on the ejecta such as radiation pressure and Keplerian shear may shorten this timescale further.  We can use the calculations presented by 
\cite{Denneau2014} (their figure 5) to estimate the likelihood  of detection. With a maximum magnitude change of $\sim2$ magnitudes
and resulting rate of dimming of $\geq0.1$ magnitudes/day, the probability of the Pan-STARRS 1 survey detecting
such a collision during the first 1.2 years  of operation was $<0.05$. Therefore we conclude that observational losses together with the rapid dispersal of the ejecta can significantly decrease the possibility of photometric detection of sub-catastrophic impacts smaller that the (596) Scheila event, and
this implies a current low probability of detection using  automated photometry software such as MOPS \citep{Denneau2013}. Finally, another additional factor may be inaccurate absolute magnitudes in current reference catalogues, as investigated by \cite{pravecbias}. They found that smaller asteroids were systematically fainter than expected from catalogues. Hence predicted magnitudes would be brighter than in reality, and any small brightness increase due to a collision could be masked by an incorrectly calculated residual magnitude.

\begin{figure}[ht]
     \begin{center}
        \subfigure[C-type target impactor, 1 day]{
            \label{fig:VCsC1}
            \includegraphics[trim = 15mm 50mm 26mm 60mm, clip, width=0.45\textwidth]{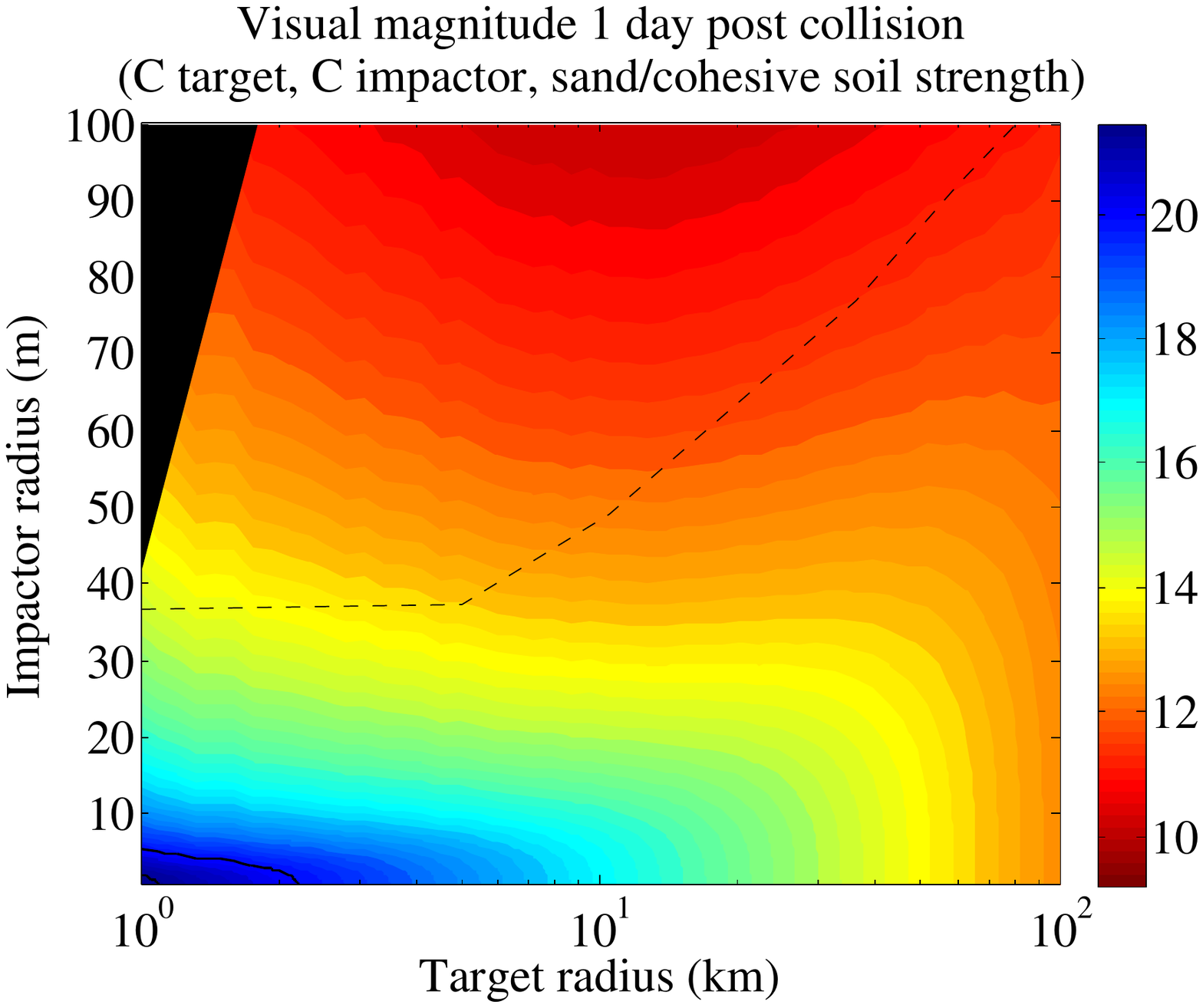}
        }
        \subfigure[C-type target impactor, 7 days]{
           \label{fig:VCsC7}
           \includegraphics[trim = 15mm 50mm 26mm 60mm, clip,width=0.45\textwidth]{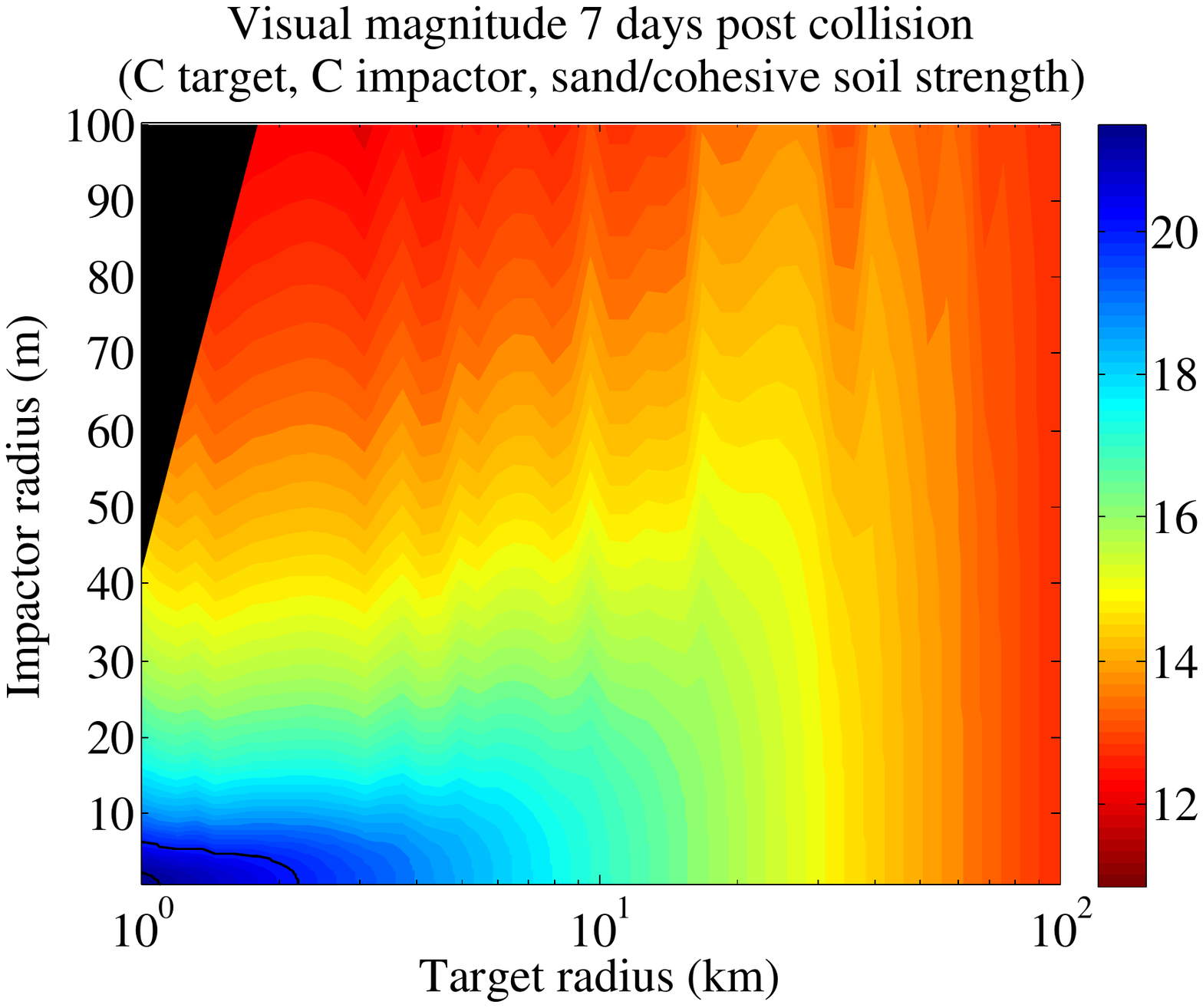}
       }
              \subfigure[S-type target impactor, 1 day]{
            \label{fig:VSsS1}
            \includegraphics[trim = 15mm 50mm 26mm 60mm, clip, width=0.45\textwidth]{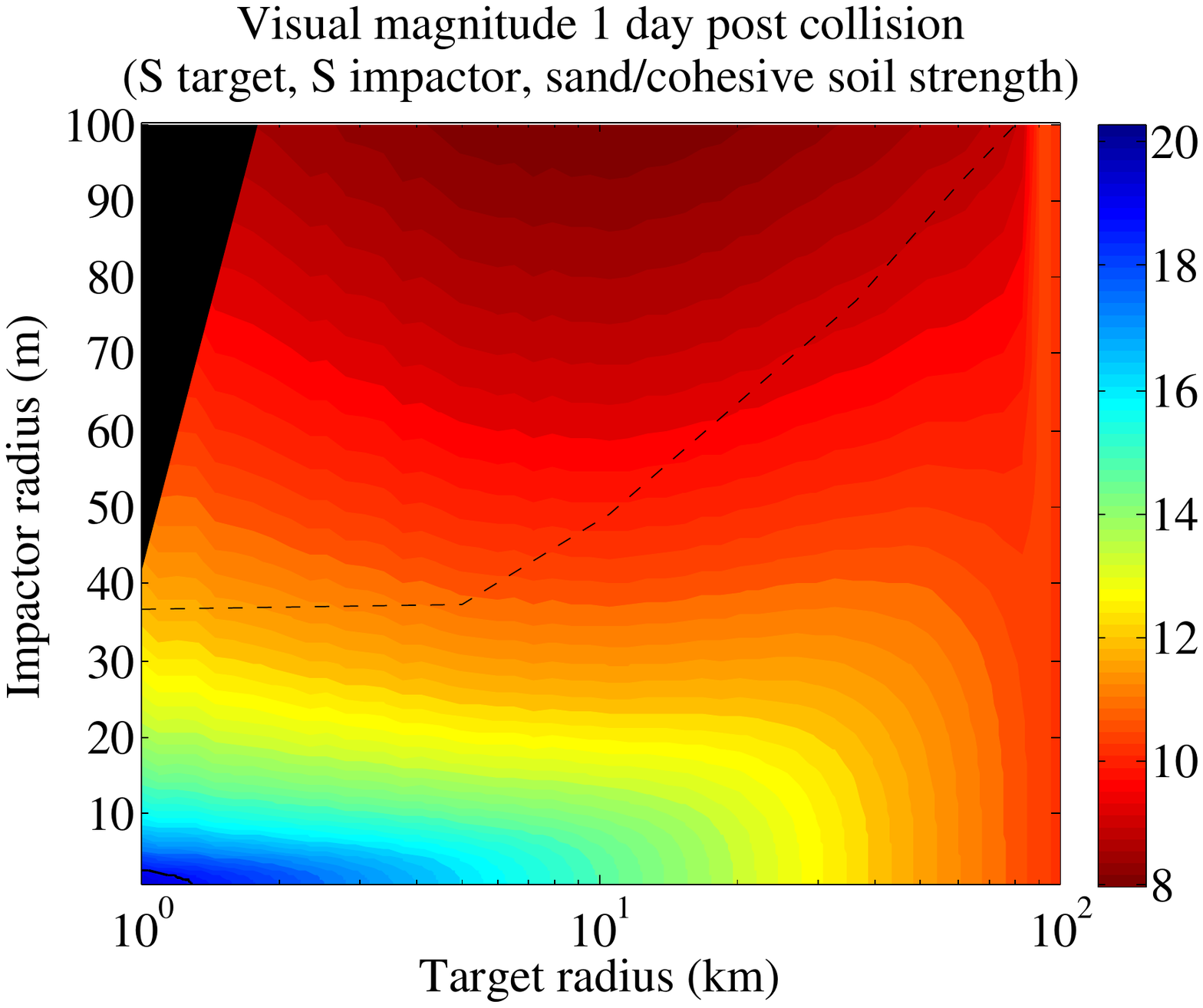}
        }
        \subfigure[S-type target impactor, 7 days]{
           \label{fig:VSsS7}
           \includegraphics[trim = 15mm 50mm 26mm 60mm, clip,width=0.45\textwidth]{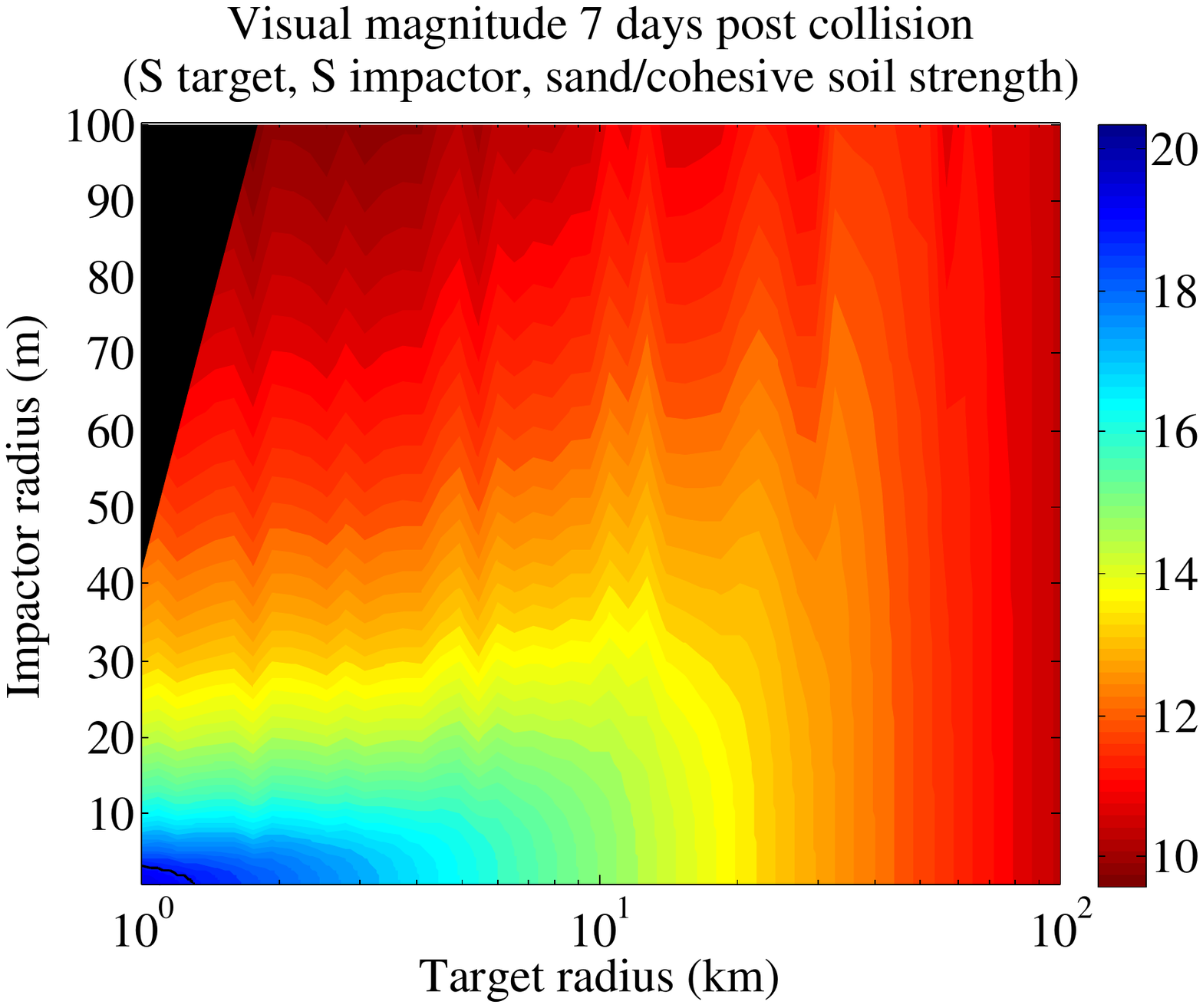}
      }

    \end{center}
    \caption{
        Visual magnitude following a collision. \ref{fig:VCsC1}-\ref{fig:VCsC7} show magnitude 1 and 7 days post-collision for C-type impactor and target in the sand/cohesive soil strength regime for a range of target and impactor radii. \ref{fig:VSsS1}-\ref{fig:VSsS7} show magnitude 1 and 7 days post-collision for S-type impactor and target in the sand/cohesive soil strength regime for a range of target and impactor radii. Impactor radii range from 1 to 100 m, target radii: 1-100 km. Colour bar shows the visual magnitude. Catastrophic disruption region is marked in black. Region where optical thickness of the debris is potentially significant is above the dashed line. }%
   \label{fig:visualmag}
\end{figure}

Of course, during the first few days to weeks the ejecta may be visible as an
extended ejecta cloud around the target asteroid, and could be detected via direct manual
observation or software algorithms designed for comet comae detection, as in the case of (596) Scheila. 
On the other hand, if the cadence of a survey is a week or less i.e. significantly less
than the decay timescale within the aperture, the chance of
photometrically detecting a small impact should be substantially higher. In Figure \ref{fig:visualmag} 
we plot the predicted total $V$-band magnitudes for C-on-C and S-on-S collisions 
1 day and 7 days after impact, for asteroids at opposition at $R_h=3.5$ AU. (Although S-types are predominantly found in the inner main belt, we calculate the predicted apparent magnitude at the likely maximum distance). 

First, it is clear that for the larger (but less frequent) impacts, the total magnitude will be relatively bright and would produce saturated images in large aperture surveys such as Pan-STARRS. However it may be possible to deal with this situation by either fitting to the wings of asteroid image point-spread function, or by recognising that an expected asteroid in the field was rejected in software processing due to its increased brightness. For fainter impact events, it is clear that
Pan-STARRS with a limiting magnitude of $V\sim22$ is able to detect such impacts throughout
the asteroid belt, as long as the asteroid is observed soon after the collision. Importantly, the forthcoming ATLAS programme  will have a nightly cadence
over the visible sky and have a limiting magnitude of $V\simeq20$ \citep{Tonry2011}. From  comparing both Figure \ref{fig:visualmag} with the brightness increases presented earlier, we find that ATLAS would  be able to detect almost all of our studied collisions as well, but here the survey cadence should not be an issue. Therefore we conclude that current and future high-cadence all sky surveys should be able to detect many more asteroid collisions at early epochs.

\clearpage

\section{Summary}
The model presented in this paper predicts the brightness increases caused by
impacts of small asteroids on larger asteroids with radii $\geq 1$ km. We use the
scaling laws of \cite{HH2007} and \cite{HH2011} to estimate
the amount of ejected material. Our model  separates the ejecta into discrete velocity shells,
and  assuming an ejecta size distribution calculates the magnitude change post-collision by calculating the surface
cross-sectional area of the asteroid plus ejected material in a small photometric aperture. The scope of the model is limited to 
non-rotating asteroids whose ejecta particles do not interact with each other, however by extending 
the model results to the large apertures used for the reported ejects from the (596) Scheila collision,
we find an estimate for the impactor size of
$40-65$ m (depending on taxonomic type) similar to previous studies in the literature.   The
model could be further improved  by introducing particle light scattering laws, radiation pressure on the ejecta and the effect of
rotation of the target asteroid. However this will lead to a significant increase the complexity of the model and will be developed in future. We believe our results
are generic enough to be used to estimate the possibility of detection of such events by current automated surveys such as Pan-STARRS1 and the Catalina Sky Survey. The model  estimates that within the parameter space examined (impactor radius $1-100$ m, target radius $1-100$ km, sub-catastrophic collisions) , a magnitude change of less than $-1$ is observable by an automated survey like Pan-STARRS 1 with effective aperture radii of 1000 km for only 10--20 days, which implies low probability of detection given the current low cadence for individual asteroids.  However,  detection may still be possible by direct manual observation or software algorithms designed for comet coma detection.

\section{Acknowledgements}

EMcL acknowledges support from the Astrophysics Research Centre, QUB. AF acknowledges support by STFC grant ST/L000709/1. EMcL and AF also thank Robert Jedicke for helpful comments during the preparation of this paper.

\clearpage

\section{Supplimentary material}

\begin{figure}[htp]
     \begin{center}
        \subfigure[1 day]{
            \label{fig:SDsand1d}
            \includegraphics[trim = {\myleft} {\mybottom} {\myright} {\mytop}, clip, width=0.45\textwidth]{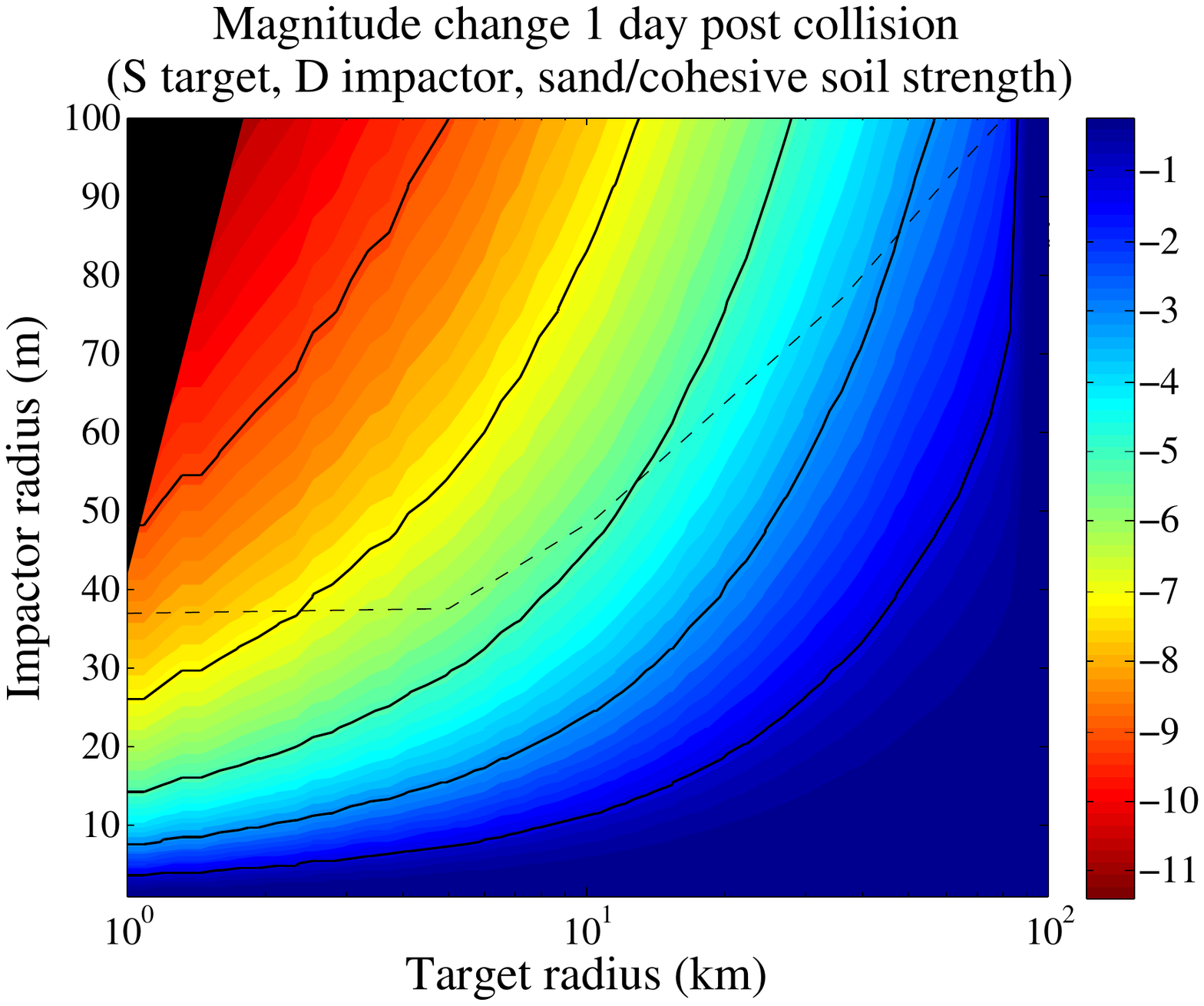}
        }
        \subfigure[ 7 days]{
           \label{fig:SDsand7d}
           \includegraphics[trim = {\myleft} {\mybottom} {\myright} {\mytop}, clip,width=0.45\textwidth]{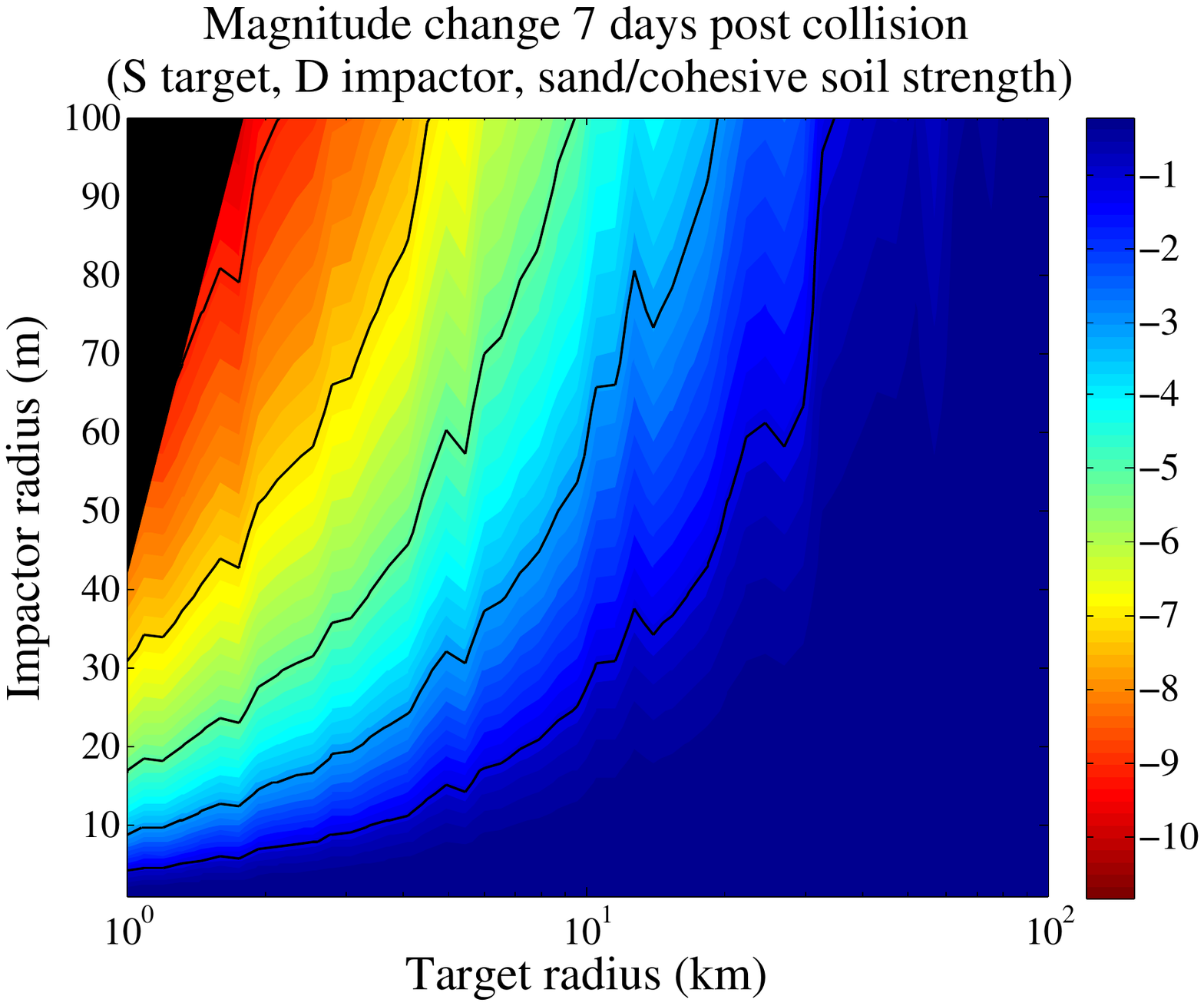}
         }

    \end{center}
    \caption{
        D-type impactor collision with S-type target, sand/cohesive soil regime: magnitude change 1 and 7 days post-collision for a range of target and impactor radii. Impactor radii range from 1 to 100 m, target radii: 1-100 km. Colour bar shows the magnitude change. Catastrophic disruption region is marked in black. Region where optical thickness of the debris is potentially significant is above the dashed line.  Solid black lines indicate contours where magnitude change is -1, -3, -5, -7 and -9.  }%
   \label{fig:subfiguresSDsand}
\end{figure}

\begin{figure}[htp]
     \begin{center}
        \subfigure[1 day]{
            \label{fig:SCsand1d}
            \includegraphics[trim = {\myleft} {\mybottom} {\myright} {\mytop}, clip, width=0.45\textwidth]{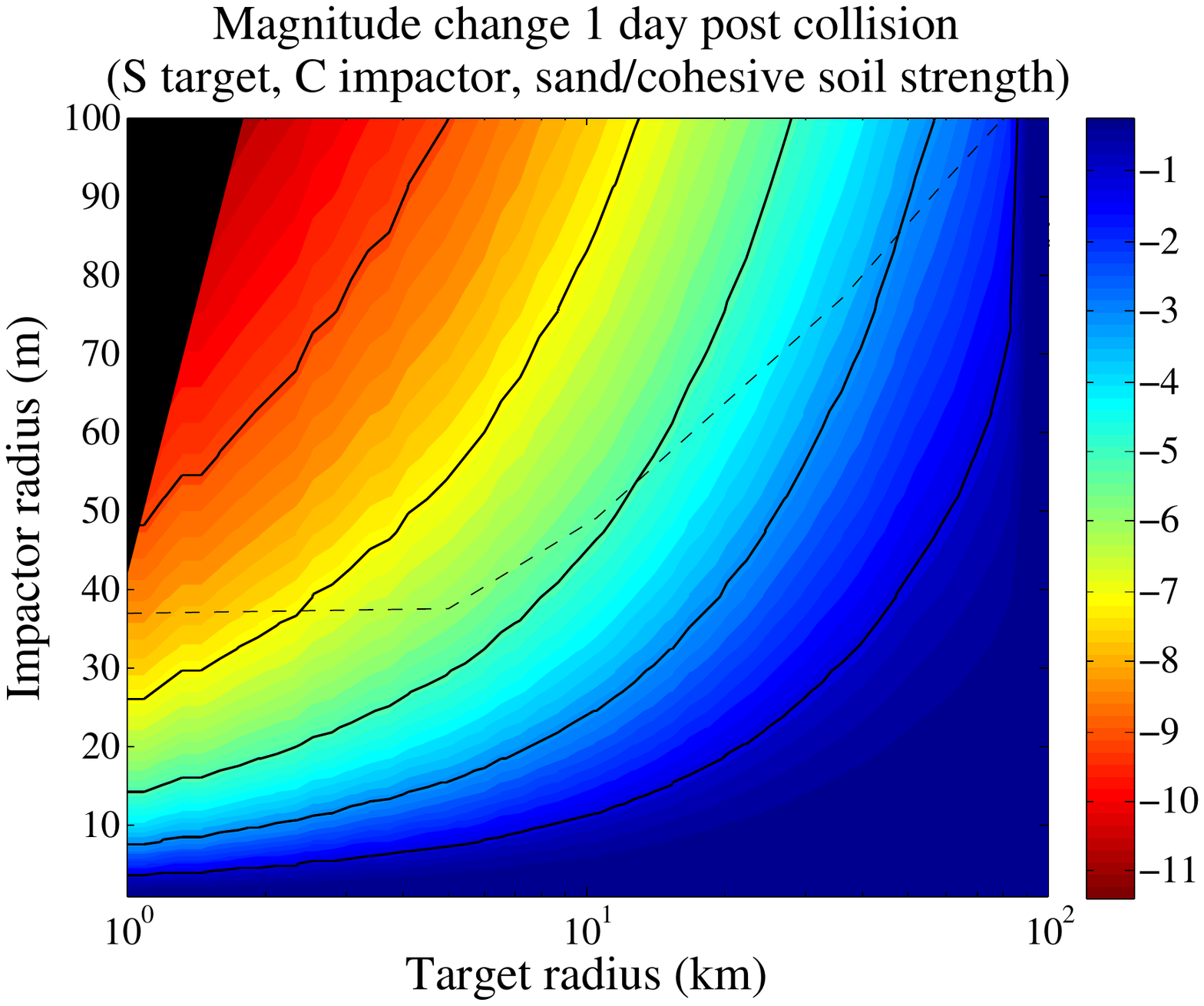}
        }
        \subfigure[ 7 days]{
           \label{fig:SCsand7d}
           \includegraphics[trim = {\myleft} {\mybottom} {\myright} {\mytop}, clip,width=0.45\textwidth]{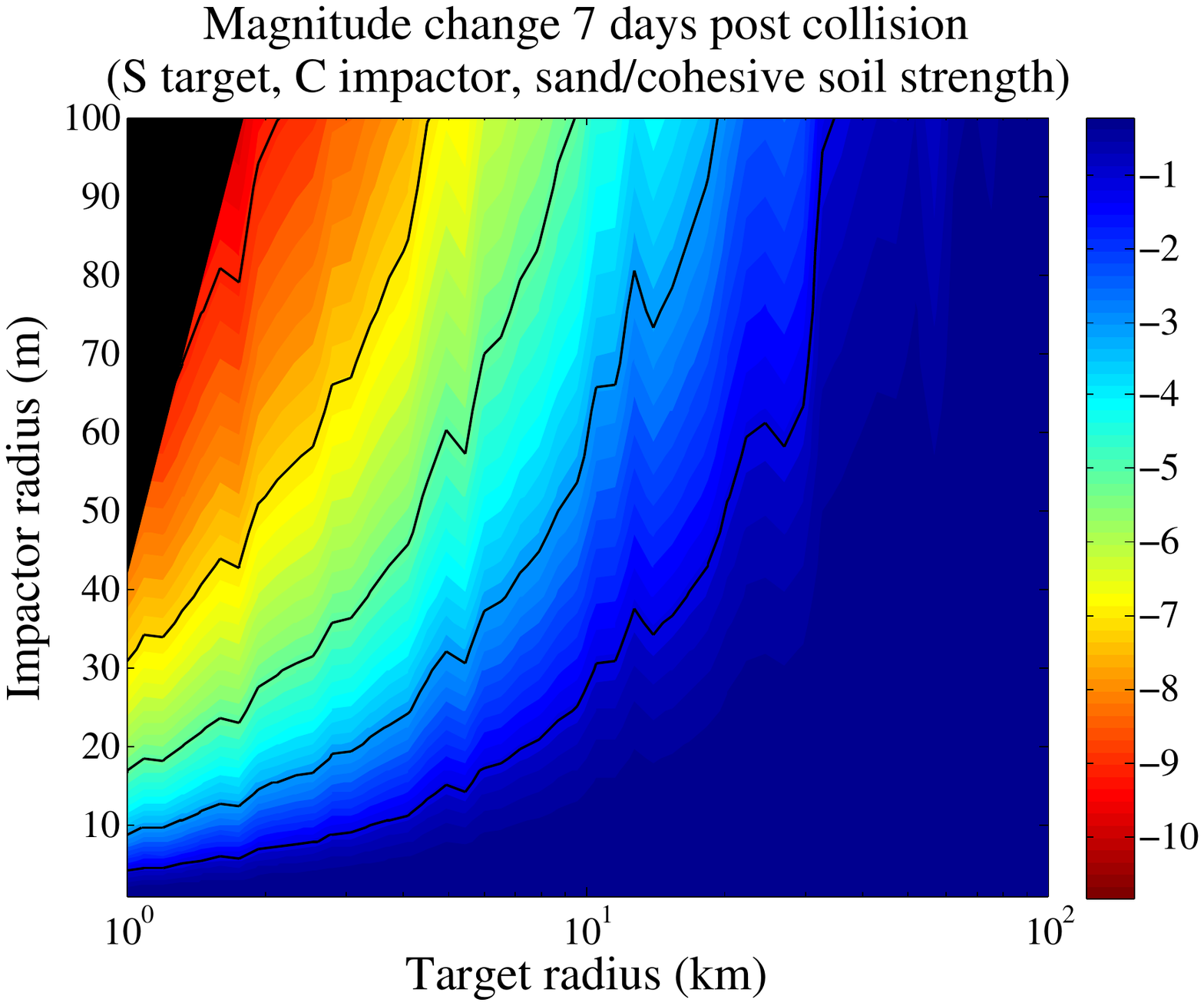}
              }

    \end{center}
    \caption{
        C-type impactor collision with S-type target, sand/cohesive soil regime: magnitude change 1 and 7 days post-collision for a range of target and impactor radii. Impactor radii range from 1 to 100 m, target radii: 1-100 km. Colour bar shows the magnitude change. Catastrophic disruption region is marked in black. Region where optical thickness of the debris is potentially significant is above the dashed line.  Solid black lines indicate contours where magnitude change is -1, -3, -5, -7 and -9.  }%
   \label{fig:subfiguresSCsand}
\end{figure}

\begin{figure}[htp]
     \begin{center}
        \subfigure[1 day]{
            \label{fig:CSsand1d}
            \includegraphics[trim = {\myleft} {\mybottom} {\myright} {\mytop}, clip, width=0.45\textwidth]{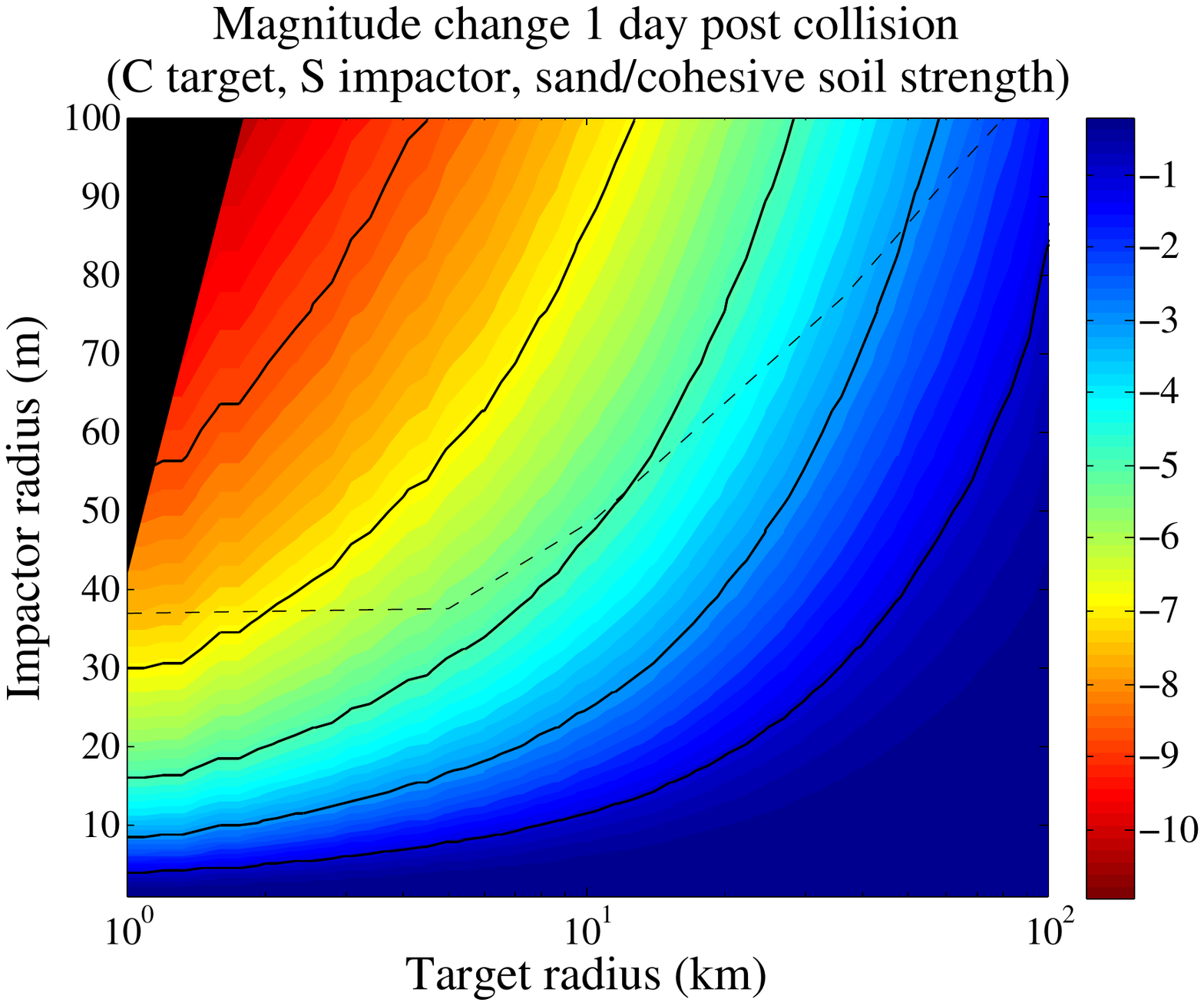}
        }
        \subfigure[ 7 days]{
           \label{fig:CSsand7d}
           \includegraphics[trim = {\myleft} {\mybottom} {\myright} {\mytop}, clip,width=0.45\textwidth]{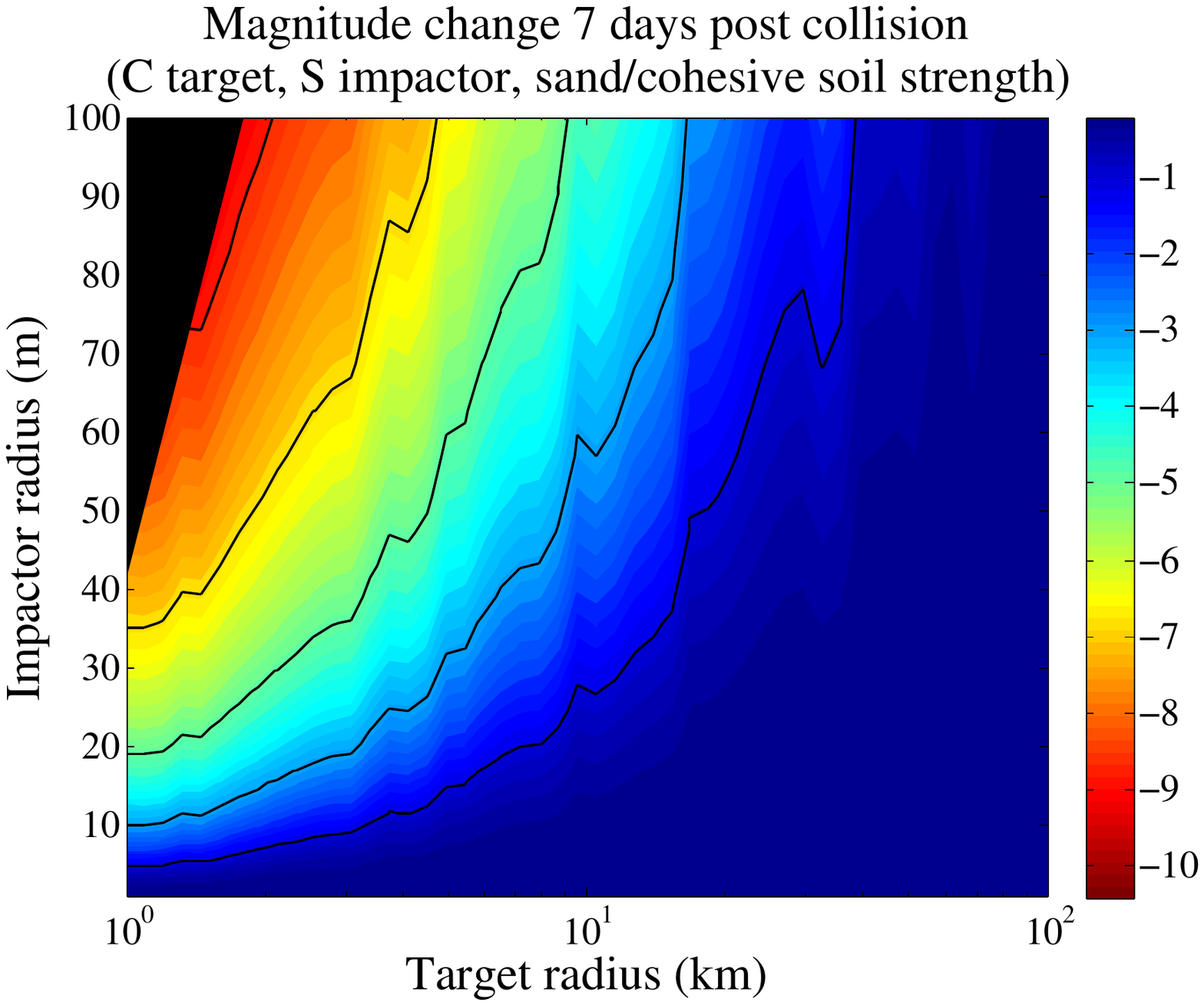}
           }
   
    \end{center}
    \caption{
        S-type impactor collision with C-type target, sand/cohesive soil regime: magnitude change 1 and 7 days post-collision for a range of target and impactor radii. Impactor radii range from 1 to 100 m, target radii: 1-100 km. Colour bar shows the magnitude change. Catastrophic disruption region is marked in black. Region where optical thickness of the debris is potentially significant is above the dashed line.   Solid black lines indicate contours where magnitude change is -1, -3, -5, -7 and -9. }%
   \label{fig:subfiguresCSsand}
\end{figure}

\begin{figure}[htp]
     \begin{center}
        \subfigure[1 day]{
            \label{fig:CDsand1d}
            \includegraphics[trim = {\myleft} {\mybottom} {\myright} {\mytop}, clip, width=0.45\textwidth]{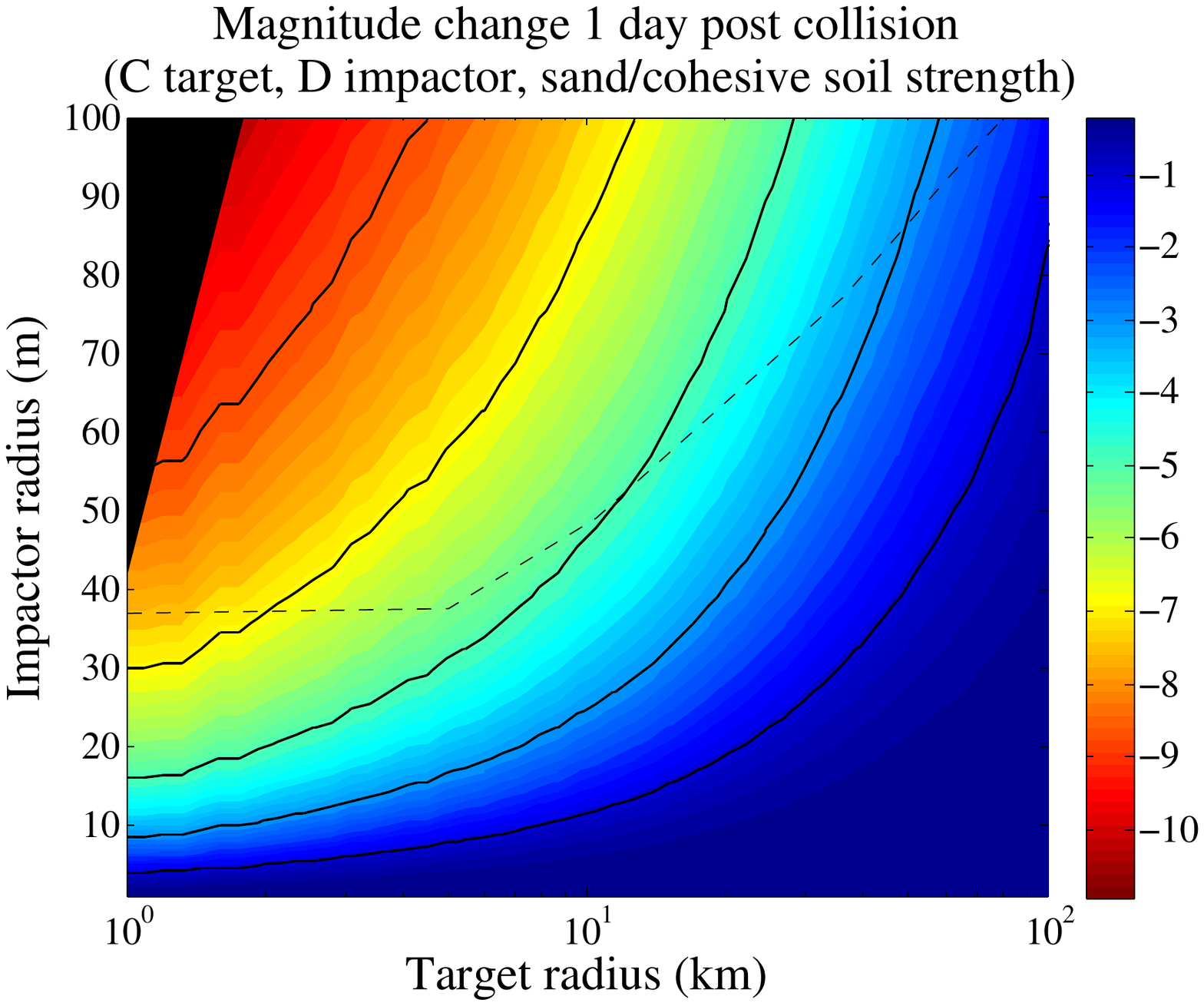}
        }
        \subfigure[ 7 days]{
           \label{fig:CDsand7d}
           \includegraphics[trim = {\myleft} {\mybottom} {\myright} {\mytop}, clip,width=0.45\textwidth]{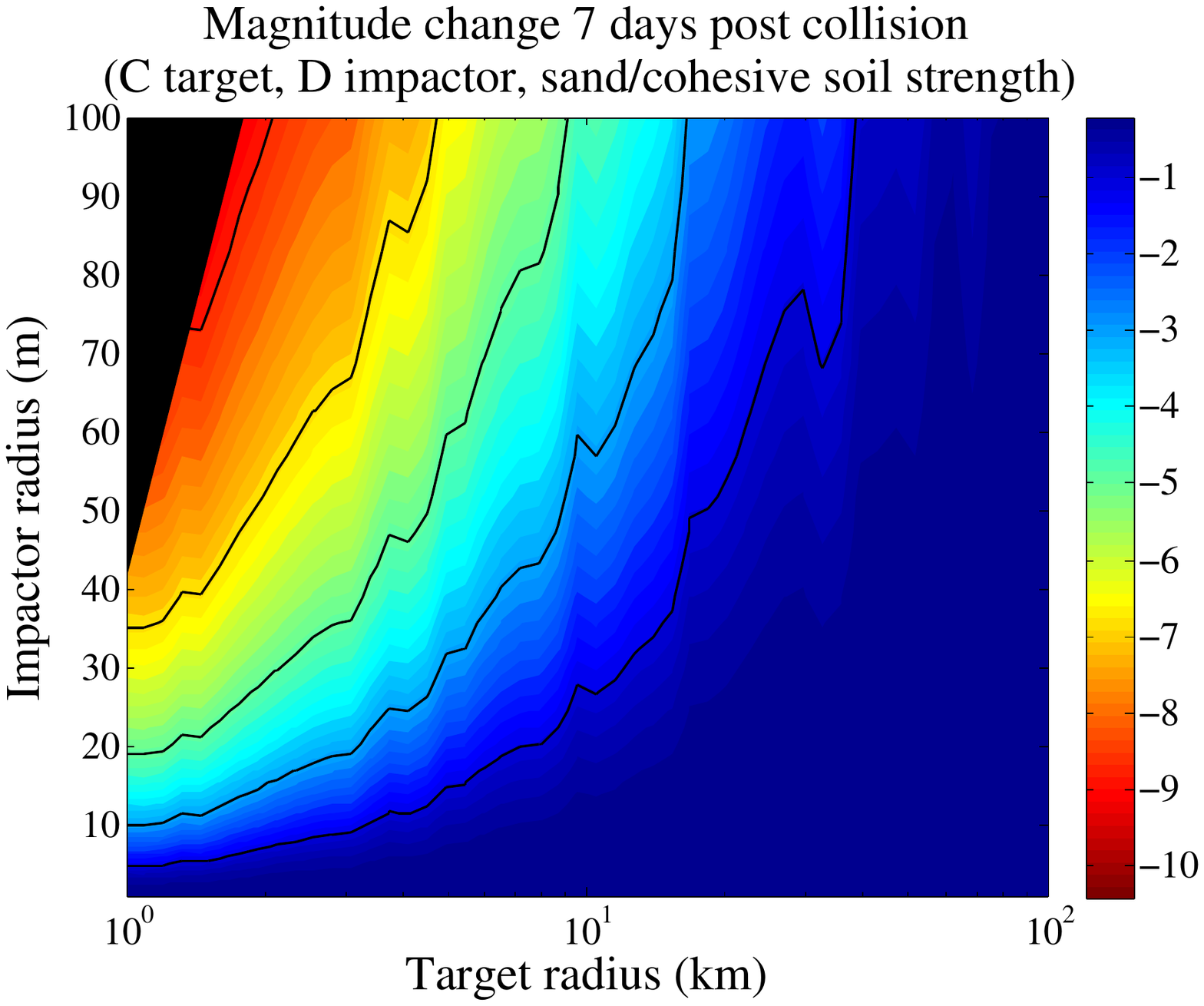}
        }

    \end{center}
    \caption{
        D-type impactor collision with C-type target, sand/cohesive soil regime: magnitude change 1 and 7 days post-collision for a range of target and impactor radii. Impactor radii range from 1 to 100 m, target radii: 1-100 km. Colour bar shows the magnitude change. Catastrophic disruption region is marked in black. Region where optical thickness of the debris is potentially significant is above the dashed line.  Solid black lines indicate contours where magnitude change is -1, -3, -5, -7 and -9.  }%
   \label{fig:subfiguresCDsand}
\end{figure}

\begin{figure}[htp]
     \begin{center}
        \subfigure[1 day]{
            \label{fig:DSsand1d}
            \includegraphics[trim = {\myleft} {\mybottom} {\myright} {\mytop}, clip, width=0.45\textwidth]{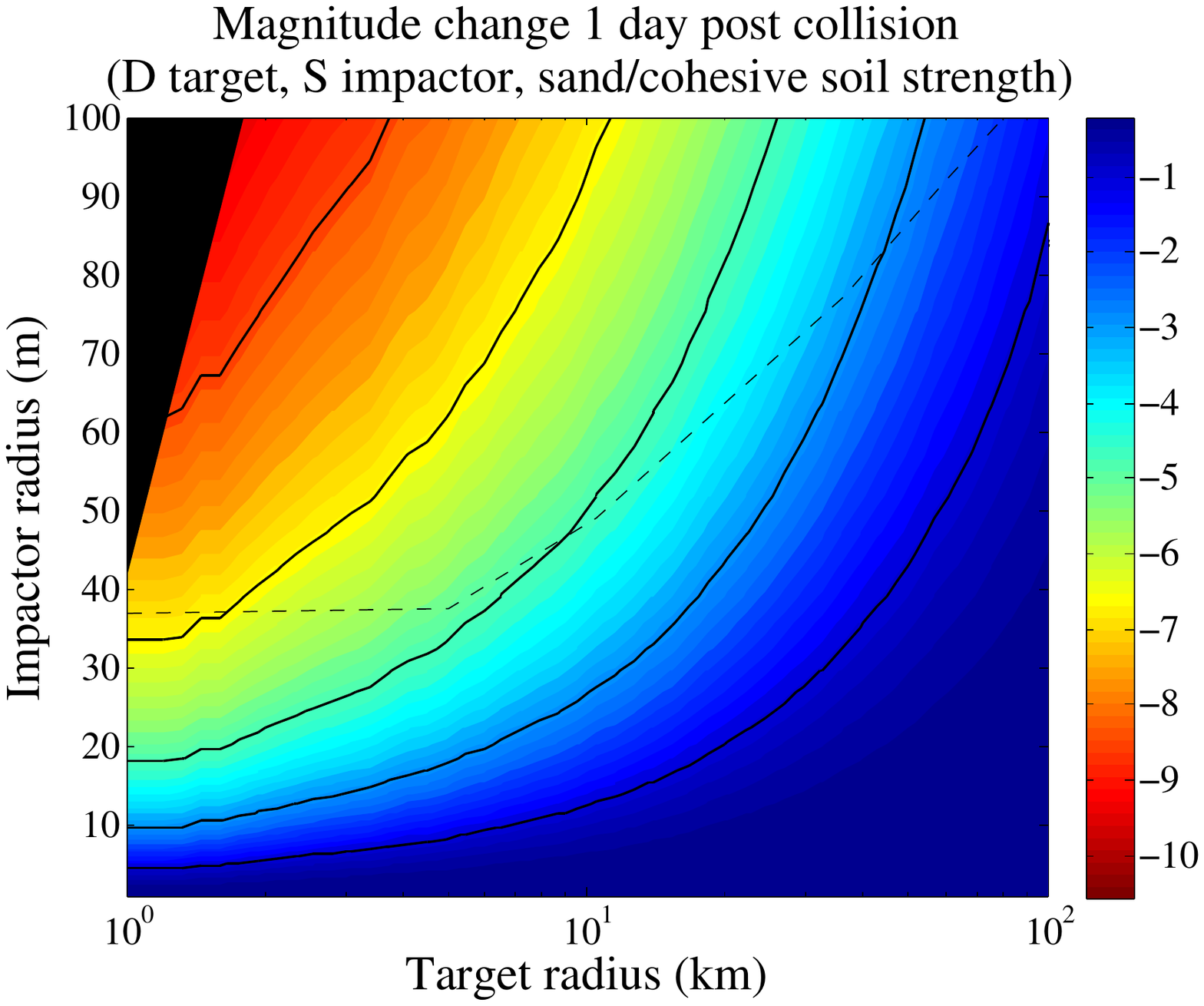}
        }
        \subfigure[ 7 days]{
           \label{fig:DSsand7d}
           \includegraphics[trim = {\myleft} {\mybottom} {\myright} {\mytop}, clip,width=0.45\textwidth]{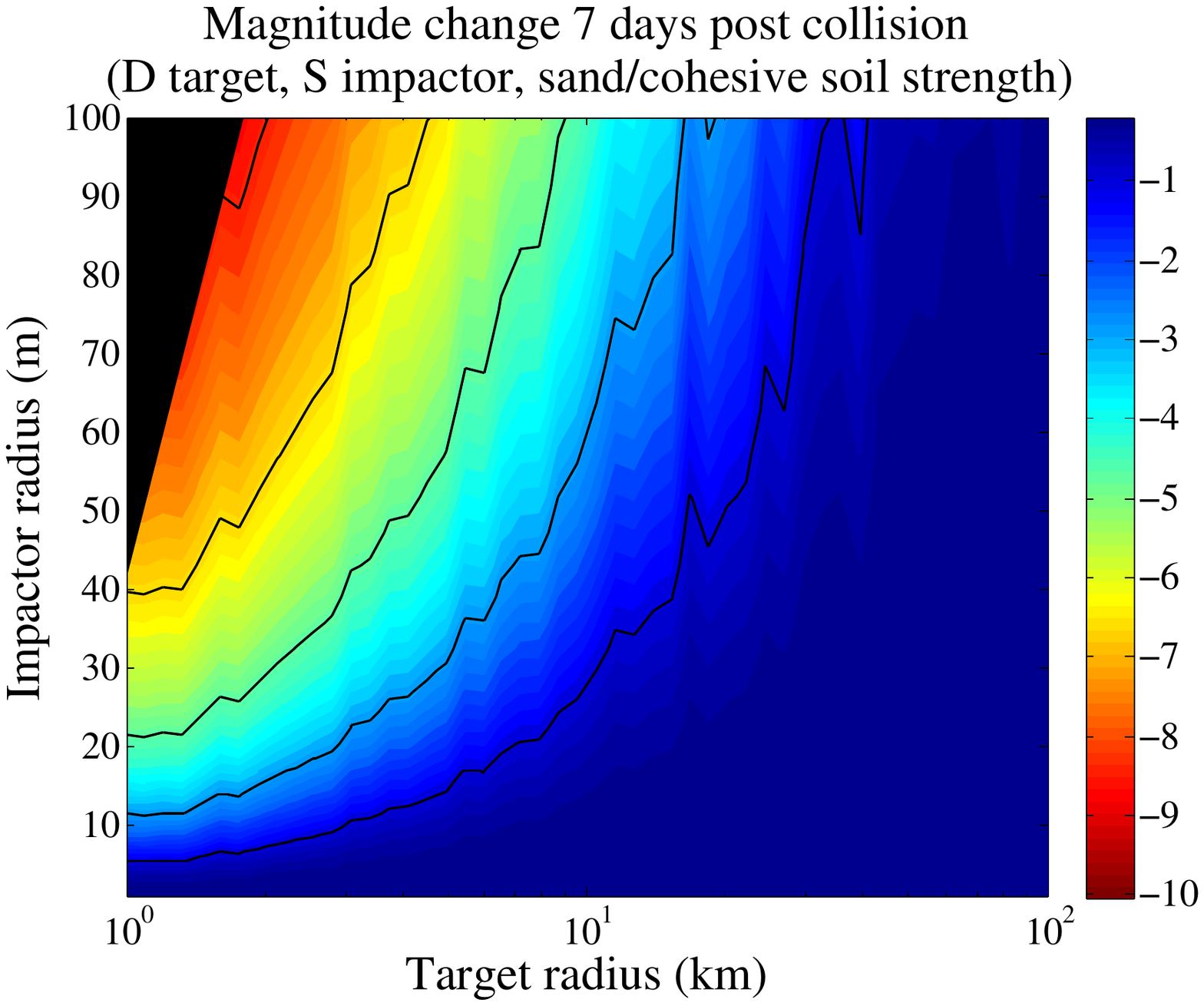}
        }

    \end{center}
    \caption{
        S-type impactor collision with D-type target, sand/cohesive soil regime: magnitude change 1 and 7 days post-collision for a range of target and impactor radii. Impactor radii range from 1 to 100 m, target radii: 1-100 km. Colour bar shows the magnitude change. Catastrophic disruption region is marked in black. Region where optical thickness of the debris is potentially significant is above the dashed line.  Solid black lines indicate contours where magnitude change is -1, -3, -5, -7 and -9.  }%
   \label{fig:subfiguresDSsand}
\end{figure}

\begin{figure}[htp]
     \begin{center}
        \subfigure[1 day]{
            \label{fig:DCsand1d}
            \includegraphics[trim = {\myleft} {\mybottom} {\myright} {\mytop}, clip, width=0.45\textwidth]{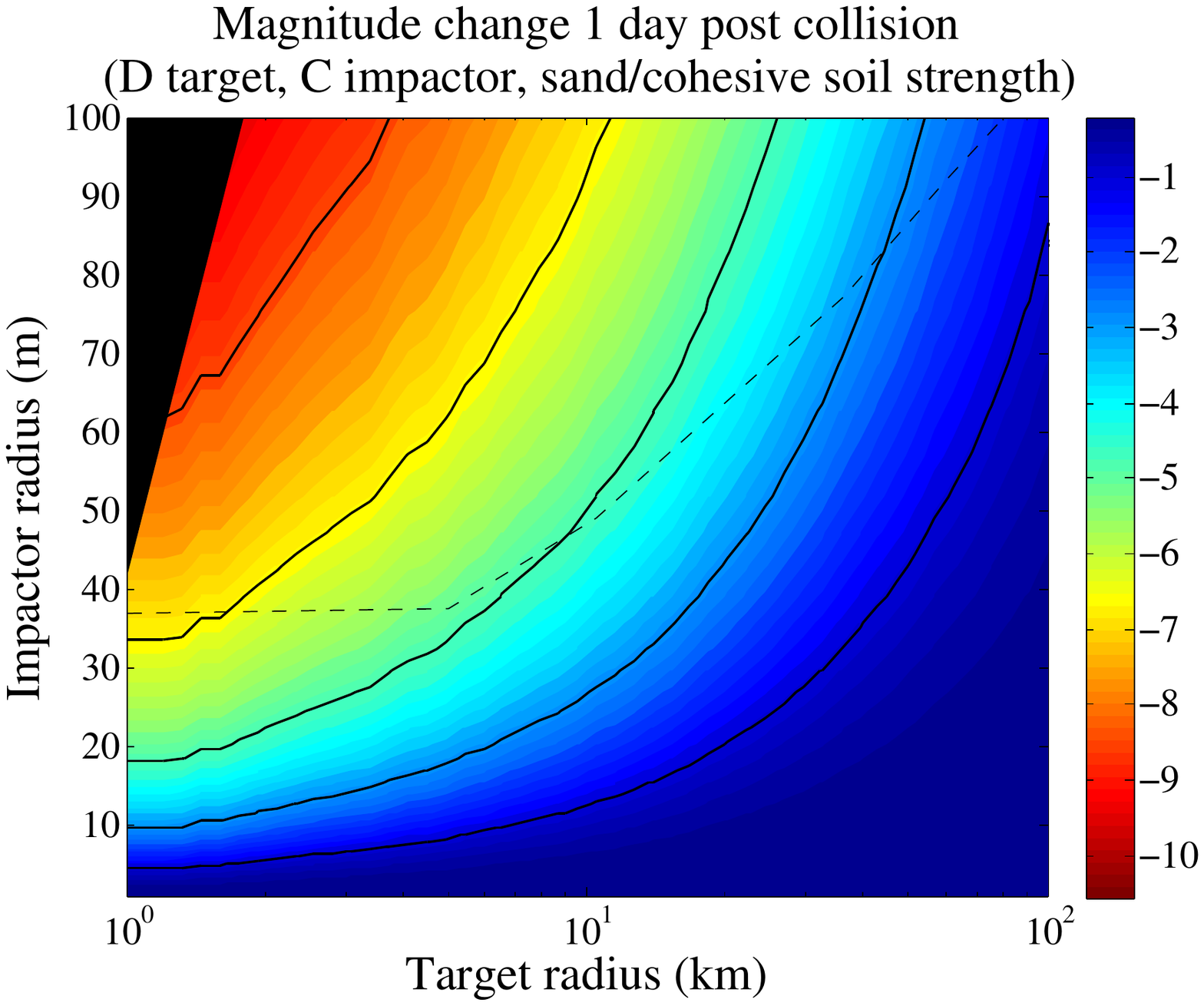}
        }
        \subfigure[ 7 days]{
           \label{fig:DCsand7d}
           \includegraphics[trim = {\myleft} {\mybottom} {\myright} {\mytop}, clip,width=0.45\textwidth]{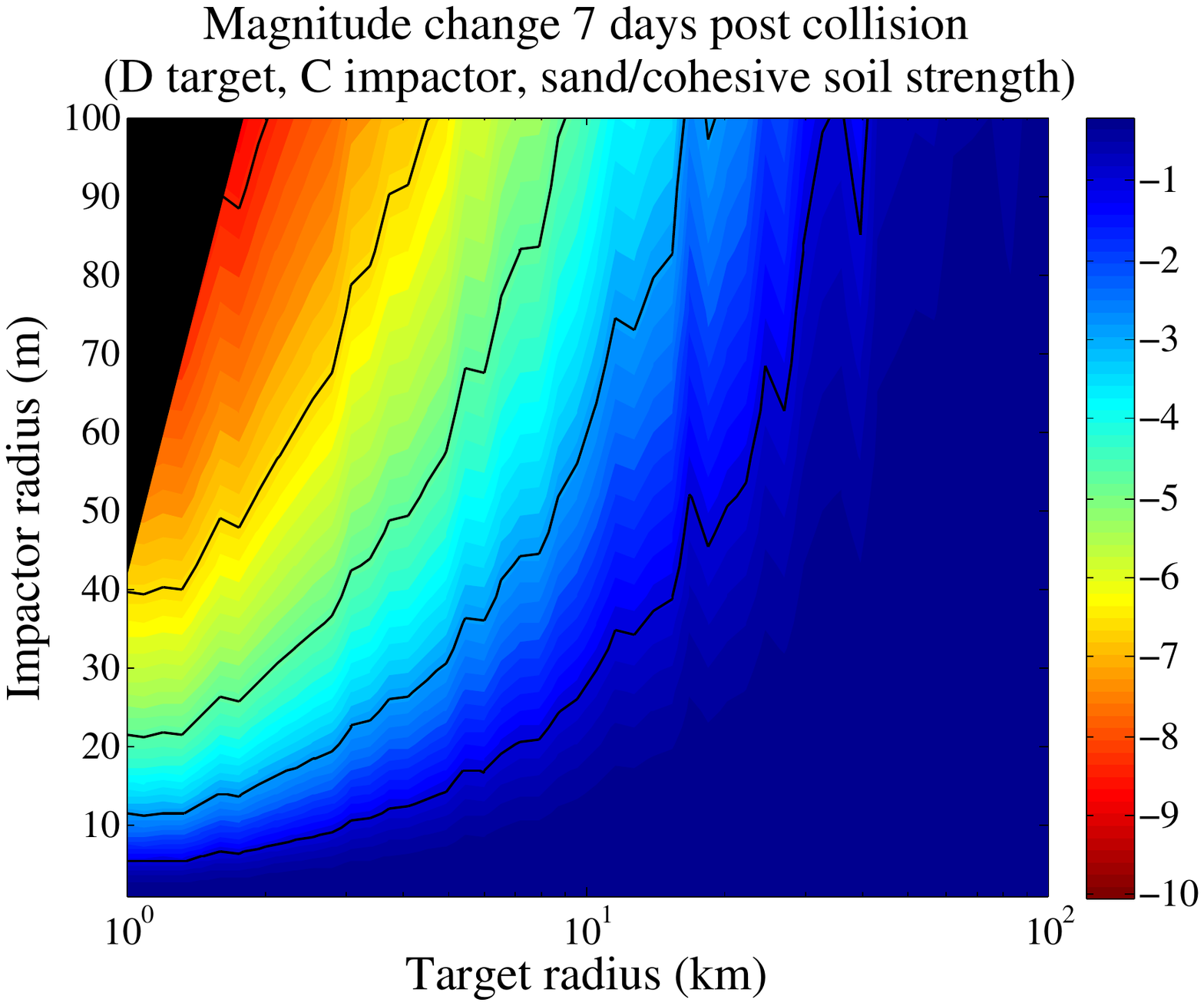}
        }

    \end{center}
    \caption{
        C-type impactor collision with D-type target, sand/cohesive soil regime: magnitude change 1 and 7 days post-collision for a range of target and impactor radii. Impactor radii range from 1 to 100 m, target radii: 1-100 km. Colour bar shows the magnitude change. Catastrophic disruption region is marked in black. Region where optical thickness of the debris is potentially significant is above the dashed line.  Solid black lines indicate contours where magnitude change is -1, -3, -5, -7 and -9.  }%
   \label{fig:subfiguresDCsand}
\end{figure}

\clearpage

\bibliography{paper_020415}

\end{document}